\documentclass[12pt,onecolumn,draftcls]{IEEEtran}
\usepackage{amsmath,amsfonts}
\usepackage{algorithmic}
\usepackage{algorithm}
\usepackage{array}
\usepackage[caption=false,font=normalsize,labelfont=sf,textfont=sf]{subfig}
\usepackage{textcomp}
\usepackage{stfloats}
\usepackage{url}
\usepackage{verbatim}
\usepackage{graphicx}
\usepackage{cite}
\usepackage{braket}
\usepackage{amssymb}
\usepackage{yhmath}
\usepackage{color}
\usepackage{orcidlink}
\usepackage{geometry}
\usepackage{setspace}
\usepackage{orcidlink}
\usepackage{hyperref}
\hypersetup{hypertex=true,
	colorlinks=true,
	linkcolor=blue,
	anchorcolor=red,citecolor=blue}
\allowdisplaybreaks[2] 

\begin{document}
	
\singlespacing

\title{Joint Beamforming Design for the STAR-RIS-Enabled ISAC Systems with Multiple Targtets and Multiple Users}

\author{Shuang Zhang, Wanming Hao,~\IEEEmembership{Senior Member,~IEEE,} Gangcan Sun, Zhengyu Zhu,~\IEEEmembership{Senior Member,~IEEE,} Xingwang Li,~\IEEEmembership{Senior Member,~IEEE,} Qingqing Wu,~\IEEEmembership{Senior Member,~IEEE} 
	\thanks{S. Zhang, W. Hao, G. Sun and Z. Zhu are with the School of Electrical and Information Engineering, Zhengzhou University, Zhengzhou 450001, China. (E-mail: yayue96@163.com, \{iewmhao,iegcsun,iezyzhu\}@zzu.edu.cn).}
	\thanks{X. Li is with the School of Physics and Electronic Information Engineering, Henan Polytechnic University, Jiaozuo 454000, China. (email: lixingwang@hpu.edu.cn).}
	\thanks{Q. Wu is with the Department of Electronic Engineering, Shanghai Jiao Tong University, Shanghai 200240, China. (email: qingqingwu@sjtu.edu.cn).}}

\maketitle

\begin{abstract}
In this paper, the sensing beam pattern gain under simultaneously transmitting and reflecting reconfigurable intelligent surfaces (STAR-RIS)-enabled integrated sensing and communications (ISAC) systems is investigated, in which multiple targets and multiple users exist. However, multiple targets detection introduces new challenges, since the STAR-RIS cannot directly send sensing beams and detect targets, the dual-functional base station (DFBS) is required to analyze the echoes of the targets. While the echoes reflected by different targets through STAR-RIS come from the same direction for the DFBS, making it impossible to distinguish them. To address the issue, we first introduce the signature sequence (SS) modulation scheme to the ISAC system, and thus, the DFBS can detect different targets by the SS-modulated sensing beams. Next, via the joint beamforming design of DFBS and STAR-RIS, we develop a max-min sensing beam pattern gain problem, and meanwhile, considering the communication quality requirements, the interference limitations of other targets and users, the passive nature constraint of STAR-RIS, and the total transmit power limitation. Then, to tackle the complex non-convex problem, we propose an alternating optimization method to divide it into two quadratic semidefinite program sub-problems and decouple the coupled variables. Drawing on mathematical transformation, semidefinite programming, as well as semidefinite relaxation techniques, these two sub-problems are iteratively sloved until convergence, and the ultimate solutions are obtained. Finally, simulation results are conducted to validate the benefits and efficiency of our proposed scheme.
\end{abstract}


\section{Introduction}
\IEEEPARstart{R}{ecently,} integrated sensing and communication (ISAC) has attracted widespread discussions as a promising enabling technology to drive sixth-generation (6G) development and become a hot topic in research \cite{ref1}. However, owing to the significant transmission path loss experienced by signals during propagation, the non-line-of-sight (NLoS) path is weak, and environmental objects easily obscure the line-of-sight (LoS) path, ISAC faces the challenges of high environmental dependence, limited coverage and high power consumption in practical applications. Especially in congested regions where path loss is predominant, the multiplicative fading of the round-trip path will cause echo signals received from the LoS path to be particularly weak or even absent \cite{ref2,ref3}. Massive multiple-input multiple-output (MIMO) techniques are commonly used to handle these concerns, but this brings high hardware costs and energy consumption \cite{ref4}.\\
\indent Fortunately, reconfigurable intelligent surfaces (RIS) have recently emerged, offering a novel approach to overcome the challenges encountered by ISAC systems \cite{ref5}. Typically, RIS consists of numerous reconfigurable elements with low power and low cost, each of which owns the capability to dynamically adjust the phase shift and amplitude of the incident signal, allowing for a controllable propagation environment \cite{ref6,ref7,ref8}. Introducing RIS into the ISAC system can bring great beneficial effects. On one hand, RIS can enlarge the coverage range \cite{ref9}. When the LoS path is blocked, constructing a virtual line of sight (VLoS) link for the system via the RIS can solve the problem of signal blind coverage and provide communication and sensing services in the blind area. When the LoS path is unobstructed, the presence of RIS can improve the signal's strength by providing an additional VLoS link. On the other hand, the deployment of RIS can enhance both communication and sensing capabilities \cite{ref10}.\\
\IEEEpubidadjcol
\indent Currently, there have been several studies endeavours aimed at investigating the incorporation of RIS into ISAC systems, in which the joint beamforming optimization of RIS and base station (BS) is proposed to improve the ISAC system performance \cite{ref11,ref12,ref13,ref14}. For example, the authors \cite{ref11} proposed to apply RIS to the dual-functional radar and communication (DFRC) system, focusing on a single-target and single-user scenario. The paper addressed the maximization of radar signal-to-noise ratio (SNR) within the limitations of transmit power and communication quality, and it introduced an iterative optimization algorithm based on the techniques of majorization-minimization (MM), semidefinite relaxation (SDR) and semidefinite programming (SDP). The authors \cite{ref12} extended their work to a multi-user setting, where they jointly designed the reflection coefficients of RIS, as well as the transmit beamforming and receive filter at BS, aiming to maximize the sum rate while ensuring reliable sensing performance. To address the optimization problem, an efficient iterative algorithm framework that leveraged the alternative direction method of multipliers (ADMM), fractional programming (FP) and MM techniques were introduced. The authors \cite{ref13} further considered the multi-target scenario, and an algorithm framework based on MM and ADMM was developed to maximize the minimum beam pattern gain of IRS toward the desired sensing angles while simultaneously satisfying the communication requirements and transmit power budget. In \cite{ref14}, a more general scenario was discussed, in which BS executed the multi-user communications and multi-target detections with the help of RIS. The authors applied the penalty-based and manifold optimization techniques, along with MM methods, to maximize the weighted sum of sensing SNRs, and meanwhile subject to the constraints of the maximum transmit power and quality-of-service (QoS). 

However, the above-mentioned works all assume that the RIS solely possess the capability to reflect incident signals, which is refered as reflection based RIS. In this case, transmitters and receivers must be deployed on the same side of the RIS, which severely limits the flexibility and efficacy of RIS deployment. Nevertheless, in practical applications, targets and users may be randomly distributed anywhere in the overall space, and thus, only serving half-space is unreasonable. Consequently, the simultaneously transmitting and reflecting reconfigurable intelligent surfaces (STAR-RISs) as an innovative concept came into being \cite{ref15,ref16,ref17,ref18}. STAR-RIS can split the incident signal into two components. One component is reflected back into the same space as the incident signal, while another is transmitted into the space opposite the incident signal. By regulating the magnetic and electric currents of the elements on STAR-RIS \cite{ref19}, it becomes feasible to independently control the phase shift and amplitude of the reflected and transmitted components, thereby achieving highly flexible space coverage. 

At present, most of the studies on STAR-RIS are focused on applying STAR-RIS to assist wireless communication systems \cite{ref20,ref21,ref22,ref23,ref24,ref25,ref26}, while the study on STAR-RIS-assisted ISAC systems is still in the initial state \cite{ref27,ref28,ref29,ref30,ref31}. To be specific, the authors \cite{ref27} first brought up the framework of the STAR-RIS-enabled ISAC system. To overcome the severe path loss and interference from clutter in sensing, a new type of STAR-RIS sensor structure was introduced, where the dedicated sensors were installed at the STAR-RIS. To minimize the sensing Cramér-Rao bound (CRB) while meeting the communication requirements, the complex minimization problem of sensing CRB was reformulated as a manipulable optimization problem involving the modified fisher information matrix, and the effectiveness of using STAR-RIS to simultaneously promote precise sensing and high-quality communication was verified. Furthermore, to explore the ability of the typical STAR-RIS in ISAC systems, the authors \cite{ref28} considered an ISAC network enhanced by energy-splitting STAR-RIS, where the dual-functional signal transmitted by a dual-functional base station (DFBS) integrated communication and detection waveforms, enabling the detection of single target with the help of STAR-RIS. Under the objective of maximizing sensing SNR, an iterative algorithm was created based on the MM, the sequential rank-one constraint relaxation and SDR techniques. The authors \cite{ref29} undertook the initial exploration of leveraging STAR-RIS to enhance the security efficacy of ISAC systems and formulated a problem aimed at maximizing secrecy rates while ensuring the sensing minimum SNR. The authors \cite{ref30} further considered the physical layer security of ISAC systems with the presence of multiple eavesdropping targets and concentrated on the average received radar sensing power maximization under QoS constraint, security requirement for eavesdroppers, and practical waveform design limitations. To tackle these issues, a low-complexity approach based on distance-majorization was conducted. And as a further advance, the authors \cite{ref31} utilized the long-term synergistic effect between STAR-RIS and ISAC and the two deep reinforcement learning algorithms to augment the mean long-term security rate of legitimate users.

Although significant performance gains have been achieved in the ISAC system with the help of RIS or STAR-RIS, most recent studies focus on single-target scenarios, and the proposed solutions may not be directly suitable for the multiple targets situation. Additionally, as STAR-RIS/RIS lacks the capability to actively transmit beams or analyze received echoes, the estimation of target arrival direction must be processed at the BS \cite{ref32}. Hence, due to the fact that the echoes of different targets reflected by STAR-RIS/RIS all pass via the same STAR-RIS/RIS-BS link, the direction of these targets is the same for BS, making it difficult to distinguish different targets \cite{ref33,ref34}.

Therefore, inspired by the above discussions and paper \cite{ref35}, this paper introduces the signature sequence (SS) modulation scheme into a STAR-RIS-enabled ISAC system with multiple targets and multiple users to overcome the difficulty that BS can not simultaneously perform multi-target detections when the received echo signals come from the same direction. And for the proposed system, the joint beamforming design of the DFBS and STAR-RIS is examined to maximize the minimum sensing beam pattern gain. The main contributions of this paper are summarized as follows:
 
\begin{itemize}
	
\item{We study a STAR-RIS-enabled ISAC system with multiple targets and multiple users, which partitions the overall space into a detection area containing multiple targets and a communication area accommodating multiple users. Then we introduce the SS modulation method into the system to overcome the dilemma of DFBS being unable to distinguish different targets. Utilizing this configuration, we analyze how DFBS performs target differentiation and detection, and then formulate an optimization problem aimed at maximizing the minimum sensing beam pattern gain through the joint optimization of DFBS and STAR-RIS beamforming. Additionally, we take into account the interference threshold, the minimum communication SNR constraint, the STAR-RIS passive constraint and the maximum total power limitations.}

\item{Since directly solving the formulated problem is difficult, we divide it into two sub-problems by fixing variables, i.e., the DFBS transmission beamforming optimization sub-problem and the STAR-RIS phase coefficients optimization sub-problem. Concretely, we first transform the formulated max-min problem into a maximization optimization problem by introducing a variable as the minimum beam pattern gain, and we rewrite several constraints to facilitate subsequent analysis and derivation. Next, the alternating optimization (AO) algorithm is used to decouple the original optimization problem into two sub-ones. Then, by defining the semi-definite auxiliary variables and relaxing the rank-one constraint via SDR, the sub-ones can be transformed into the standard convex quadratic semi-definite program (QSDP) problem and solved through SDP. After obtaining the solution, Gaussian randomization or eigenvalue decomposition techniques can be employed to approximate the solution of the relaxed rank-one constraint.} 

\item{Finally, the simulation results demonstrate the advantages of deploying STAR-RIS in the ISAC system and validate that our proposed scheme can obtain a higher performance compared to other baseline schemes. Our study also indicates that as the restrictions of interference become smaller, the beam pattern gain will decrease. Therefore, interference restrictions are necessary for DFBS to detect targets more accurately.} 
\end{itemize}

\textit{Notations}: Throughout this paper, the following notations are employed. Column vectors are typically described by boldface lower-case letters, whereas matrices are commonly described by boldface upper-case letters.
$(\cdot)^{T}$, $(\cdot)^{H}$, and $(\cdot)^{-1}$ indicate the transpose, conjugate-transpose and the inversion operations of a matrix, respectively. For a matrix $\boldsymbol{V}, \operatorname{Rank}(\boldsymbol{V}), \operatorname{Tr}(\boldsymbol{V})$, respectively, indicate the rank and the trace of $\boldsymbol{V}$, $\boldsymbol{V} \succeq 0$ denotes $\boldsymbol{V}$ is a positive semidefinite matrix, $\|\boldsymbol{V}\|_{F}$ indicates the F-norm of $\boldsymbol{V}$, diag$(\boldsymbol{V})$ refers to extracting the diagonal elements from $\boldsymbol{V}$. For a complex variable $v$, $|v|$ refers to its modulus, diag$(v)$ refers to forming a diagonal matrix based on $v$.
$\mathbb{E\{\cdot\}}$ denotes the statistical expectation vaule. $\mathbb{C}^{m\times n}$ and $\mathbb{R}^{m\times n}$ indicate the complex space and the real space with a dimension of $m\times n$, respectively. In addition, $\mathbf{I}_{N}$ indicates the identity matrix with dimension $N\times N$.  

\section{System Model and Problem Formulation}
\begin{figure*}[ht]
	\centering
	\includegraphics[height=3.6in,width=6.5in]{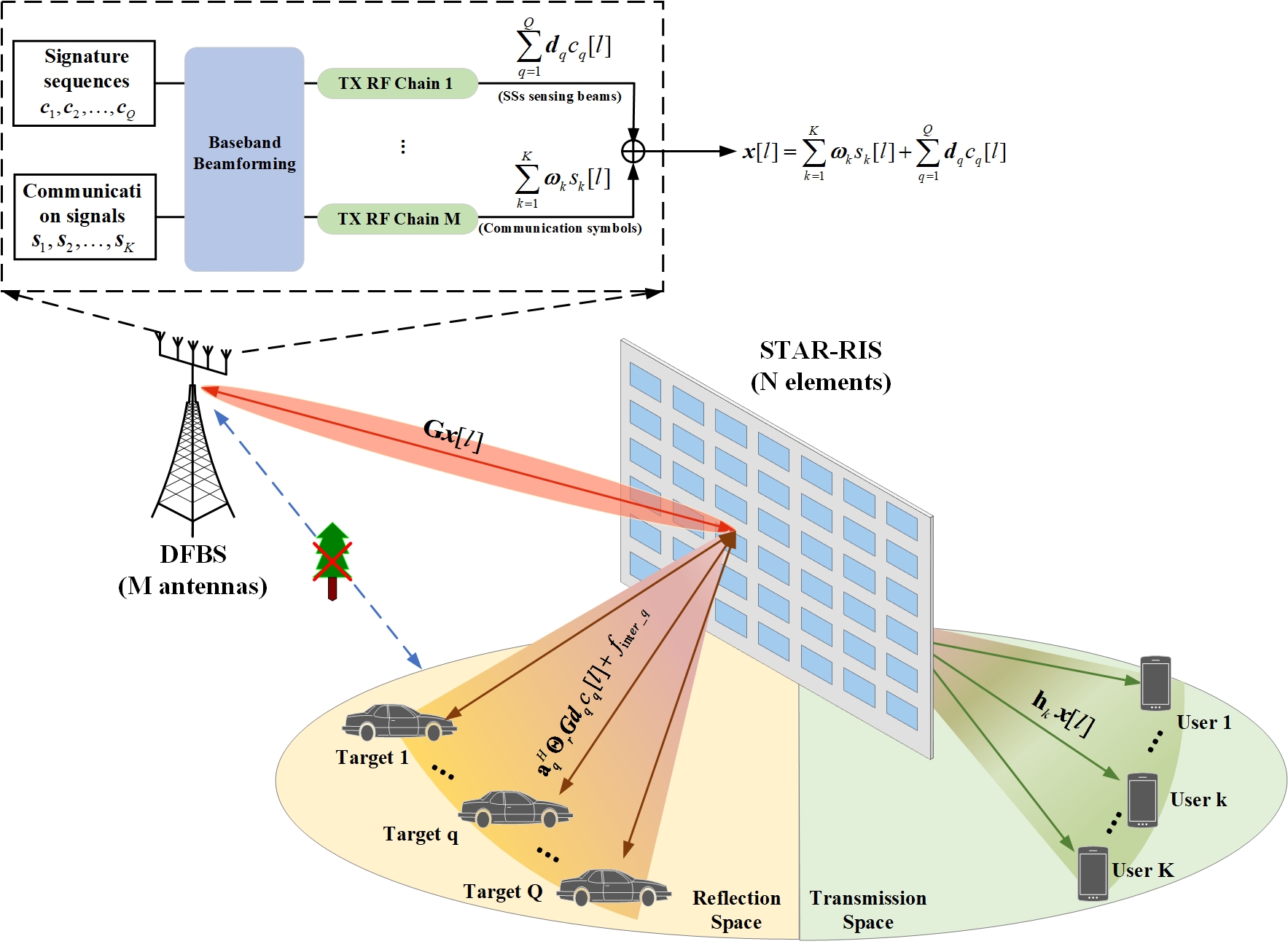}
	\hfil
	\caption{STAR-RIS-enabled ISAC system model.}
	\label{fig_1}
\end{figure*}
In this section, STAR-RIS-enabled ISAC system with multiple targets and multiple users is first described. Then, a max-min fairness optimization problem of sensing beam pattern gain is formulated.

\subsection{System Model}

We consider an ISAC system, in which the DFBS simultaneously detects multiple targets and communicates with multiple users under the assistance of STAR-RIS. There are $M$ antennas denoted by the set of $\mathcal{M}$ that are arranged as uniform linear arrays (ULAs) at the DFBS, and $N$ transmission \& reflection elements at the STAR-RIS denoted by the set of $\mathcal{N}$ that are deployed in the form of uniform planar arrays (UPAs). As depicted in Fig. \ref{fig_1}, the entire space is split into two distinct half-ones, i.e., the reflection space and the transmission space. Without loss of generality, we assume that all $Q$ sensing targets denoted by the set of $\mathcal Q$ are situated at the reflection space, and all $K$ communication users with single-antenna denoted by the set of $\mathcal K$ are located at the transmission space. Additionally, it is assumed that there exist obstacles between the DFBS and targets \& users, and thus the LoS links are blocked.

Let $\boldsymbol{\Phi}_{t}\in\mathbb{C}^{N\times N}$ and $\boldsymbol{\Phi}_{r}\in\mathbb{C}^{N\times N}$, respectively, represent the transmission coefficients (TCs) matrix and reflection coefficients (RCs) matrix of STAR-RIS, which can be mathematically expressed by:

\begin{equation}
	\label{eqn_1}
	\boldsymbol{\Phi}_{i}=\text{diag}(\beta_{i,1}e^{j\boldsymbol{\phi}_{i,1}},\beta_{i,2}e^{j\boldsymbol{\phi}_{i,2}},\dots,\beta_{i,N}e^{j\boldsymbol{\phi}_{i,N}}),
\end{equation}
where, $\boldsymbol{\phi}_{i,n}\in(0,2\pi]$ and $\beta_{i,n}\in[0,1]$, respectively, denote the phase-shift coefficient and amplitude coefficient of RCs($i=r$)/TCs($i=t$) of the $n$-th STAR-RIS element. For each STAR-RIS element $n, \forall n \in \mathcal{N}$, the phase-shift coefficients $\{{\phi}_{r,n}, {\phi}_{t,n}\}$ can be adjusted independently, while the amplitude coefficients $\{\beta_{r,n},\beta_{t,n}\}$ are constrained by the law of energy conservation, resulting in a coupling relationship between them, i.e., $\beta_{r,n}^2+\beta_{t,n}^2=1$.

\subsection{Downlink Transimit Signal}

In order to carry out communication and detection simultaneously, the DFBS transmits the dual-functional signal $\boldsymbol{x}[l]\in\mathbb{C}^{M\times 1}$ at the time slot $l$, which is a combination of dedicated sensing beams and communication symbols: 

\begin{equation}
	\label{eqn_2}
	\boldsymbol{x}[l]=\sum_{k=1}^K \boldsymbol{\omega}_k s_k[l]+\sum_{q=1}^Q \boldsymbol{d}_q c_q[l].
\end{equation}

According to \cite{ref35}, we assume that the dedicated sensing beams are modulated by SSs $\mathbf{c}_q=\left[c_q[1], \ldots, c_q[L_p]\right]^T \in \mathbb{C}^{L_p \times 1}, 
\forall q\in\mathcal{Q}$, where $c_q[l]$ is the SS for detecting the $q$-th target at the time slot $l$, and $L_p$ is the number of pulses. $d_q \in \mathbb{C}^{M \times 1}$ is the corresponding sensing beamforming for the $q$-th target. In the SS scheme, the DFBS can distinguish different targets from the echoes based on the orthogonal codebook, i.e., $\mathbf{c}_q^H \mathbf{c}_q=L_p, \mathbf{c}_q^H \mathbf{c}_q=0$ and $\left|c_q[l]\right|^2=1$, where $q \neq q^{\prime}, \forall q, q^{\prime} \in \mathcal{Q}$. In addition, $\mathbf{s}_k=\left[s_k[1], \ldots, s_k[L_p]\right]^T \in \mathbb{C}^{L_p \times 1}$ is the information-bearing symbols for the $k$-th users and $s_k[l]$ is the symbol at the time slot $l$ for the $k$-th user, $\omega_k \in \mathbb{C}^{M \times 1}$ is the corresponding communication beamforming for the $k$-th user. Similarly, to avoid mutual interference in sensing and communication, the communication symbols and sensing beams SSs are considered to be mutually independent, and the communication symbols for distinct users are also assumed to be uncorrelated, i.e., $s_k[l] \sim \mathcal C\mathcal N(0,1)$, $\mathbb{E}\left\{s_k[l] s_{k^{\prime}}^H[l]\right\}=0$, where $k \neq k^{\prime}, \forall k, k^{\prime} \in \mathcal{K}$ and $\mathbb{E}\left\{c_q[l] s_k^*[l]\right\}=0, \forall q \in \mathcal{Q}, \forall k \in \mathcal{K}$.

\subsection{Radar Sensing Model}

By exploiting the SS modulation, as elaborated in [35], the desired direction $\left\{\theta_q^{azi}, \theta_q^{ele}\right\}$ and SS $\mathbf{c}_q$ of the $q$-th target is a one-to-one mapping, where $\theta_q^{azi}$ and $\theta_q^{ele}$, respectively, denote the azimuth and elevation angles of the $q$-th target relative to the STAR-RIS. Thus, the sensing beam pattern gain of STAR-RIS in direction $\left\{\theta_q^{azi}, \theta_q^{ele}\right\}$ for the $q$-th target can be given by:

\begin{equation}
	\label{eqn_3}
	\mathcal{P}_q=\mathbb{E}\left(\left|\mathbf{a}_q^H \boldsymbol{\Phi}_\tau \mathbf{G} \boldsymbol{d}_q c_q[l]\right|^2\right)=\mathbf{a}_q^H \boldsymbol{\Phi}_r \mathbf{G} \boldsymbol{d}_q \boldsymbol{d}_q^H \mathbf{G}^H \boldsymbol{\Phi}_r^H \mathbf{a}_q,
\end{equation}

\noindent where $\mathbf{a}_q \in \mathbb{C}^{N \times 1}$ is the steering vector of the STAR-RIS towards the $q$-th target, $\mathbf{G} \in \mathbb{C}^{N \times M}$ represents the channel from the DFBS to the STAR-RIS. We assume that STAR-RIS with $N=N_x \times N_z$ elements, where $N_x$ and $N_z$, respectively, represent the number of elements arranged along $\mathrm{x}$-axis and $\mathrm{z}$-axis, that is equivalent to deploy STAR-RIS in the $(\mathrm{X}, \mathrm{Z})$ plane. According to [27], the steering vector of STAR-RIS under UPA can be modeled as \eqref{eqn_4}, which is presented at the top of the next page. The rows in $\left[\mathbf{r}_X, \mathbf{r}_Y, \mathbf{r}_Z\right] \in \mathbb{R}^{N \times 3}$ represent the three-dimensional (3D) Cartesian coordinates of the elements in STAR-RIS. Thus for $(\mathrm{X}, \mathrm{Z})$ plane, there is $\mathbf{r}_{\mathrm{Y}}=0$ and we have $\mathbf{a}_q\left(\theta_q^{azi}, \theta_q^{ele}\right)=\exp \left(-j \frac{2 \pi d}{\lambda}\left(\mathbf{r}_X \cos \theta_q^{azi} \cos \theta_q^{ele}+\mathbf{r}_Z \sin \theta_q^{ele \epsilon}\right)\right)$. $\lambda$ is the wavelength, and $d=\lambda / 2$ denotes the antenna space. 

The received echo signal $\mathbf{y}[l] \in \mathbb{C}^{M \times 1}$ for the DFBS at the time slot $l$ can be given by \eqref{eqn_5} at the top of this page, where the first term in \eqref{eqn_5} includes the desired SS signals $c_q$ reflected from target $q$ and the interference reflected form communication users, the second term refers to the interference reflected form unassociated targets, and the final term $\mathbf{n}[l] \sim \mathcal{C N}\left(0, \sigma_z^2 \mathbf{I}_M\right)$ represents the complex additive white Gaussian noise (AWGN). In \eqref{eqn_5}, the complex channel gain $\beta_q \in \mathbb{C}$ is calculated by taking into account the round-trip path-loss and the complex reflection factor of the $q$-th target. For simplicity, it is assumed that the round-trip propagation delay, denoted as $\tau_q$, for the echo signals reflected by the $q$-th target is an integer value.

From \eqref{eqn_5}, it can be seen that in addition to the desired echo signals of each target, there are also interferences caused by the reflection signals from unexpected directions and communication signals. The interferences can lead to erroneous judgments during DFBS detection, resulting in poor sensing performance. Therefore, we introduce the interference constraints \eqref{eqn_6} at the top of this page, and $\eta$ is the interference threshold. Next, we extract the expected target signals, and rewrite \eqref{eqn_5} as:

\begin{figure*}[ht] %hb代表放在文章底部，%ht为放在文章顶部 
	\centering
	\begin{equation}
		\label{eqn_4}
		\mathbf{a}_q\left(\theta_q^{azi}, \theta_q^{ele}\right)=\exp \left(-j \frac{2 \pi d}{\lambda}\left(\mathbf{r}_X \cos \theta_q^{azi} \cos \theta_q^{ele}+\mathbf{r}_Y \sin \theta_q^{azi} \cos \theta_q^{ele}+\mathbf{r}_Z \sin \theta_q^{ele \epsilon}\right)\right) \in \mathbb{C}^{N \times 1},
	\end{equation}
\end{figure*}

\begin{figure*}[ht] %hb代表放在文章底部，%ht为放在文章顶部 
	\centering
	\begin{equation}
		\label{eqn_5}
		\begin{aligned}
			 \mathbf{y}[l]&=\sum_{q=1}^Q \mathbf{G}^H \boldsymbol{\Phi}_r^H \mathbf{a}_q \beta_q \mathbf{a}_q^H \boldsymbol{\Phi} \mathbf{G}\left(\boldsymbol{d}_q c_q\left[l-\tau_q\right]+\sum_{k=1}^K \boldsymbol{\omega}_k s_k\left[l-\tau_q\right]\right) \\
			& +\sum_{q=1}^Q \sum_{q^{\prime} \neq q} \mathbf{G}^H \boldsymbol{\Phi}_r^H \mathbf{a}_{q^{\prime}} \beta_{q^{\prime}} \mathbf{a}_{q^{\prime}}^H \boldsymbol{\Phi}_r \mathbf{G}\boldsymbol{d}_q c_q\left[l-\tau_{q^{\prime}}\right]+\mathbf{n}[l],
		\end{aligned}
	\end{equation}
\end{figure*}

\begin{figure*}[ht] %hb代表放在文章底部，%ht为放在文章顶部 
	\centering
	\begin{equation}
		\label{eqn_6}
		f_{\text {inter}}=\mathbb{E}\left(\left|\mathbf{a}_q^H \boldsymbol{\Phi}_r \mathbf{G} \sum_{k=1}^K \boldsymbol{\omega}_{k} s_k\right|^2\right)+\sum_{q^{\prime} \neq q} \mathbb{E}\left(\left|\mathbf{a}_{q^{\prime}}^H \boldsymbol{\Phi}_r \mathbf{G}\boldsymbol{d}_q c_q\right|^2\right) \leq \eta, \forall q \in \mathcal{Q}, 
	\end{equation}
\end{figure*}

\begin{equation}
	\label{eqn_7}
	\mathbf{y}[l]=\sum_{q=1}^Q \mathbf{G}^H \boldsymbol{\Phi}_r^H \mathbf{a}_q \beta_q \mathbf{a}_q^H \boldsymbol{\Phi}_r \mathbf{G} \boldsymbol{d}_q c_q\left[l-\tau_q\right]+\tilde{\mathbf{n}}[l],
\end{equation}
where $\tilde{\mathbf{n}}[l]$ represents all the interferences plus noise.

From the perspective of radar sensing, to obtain an enhanced target sensing performance, one common approach is to maximize the signal power directed towards the targets while minimizing it in other directions. In this case, DFBS needs to perform matched filtering of the received echo signal, and the optimal normalized matched filter for the $q$-th target is given by $\mathbf{u}_q^H=\frac{\left(\mathbf{G}^H \boldsymbol{\Phi}_r^H \mathbf{a}_q \beta_q \mathbf{a}_q^H \boldsymbol{\Phi}_r \mathbf{G} \boldsymbol{d}_q\right)^H}{\left\|\left(\mathbf{G}^H \boldsymbol{\Phi}_r^H \mathbf{a}_q \beta_q \mathbf{a}_q^H \boldsymbol{\Phi}_r \mathbf{G} \boldsymbol{d}_q\right)\right\|} \in \mathbb{C}^{1 \times M}$, and the output signal after being filtered can be expressed as:

\begin{equation}
	\label{eqn_8}
	\tilde{y}_q[l]=\mathbf{u}_q^H\left(\sum_{i=1}^Q \mathbf{G}^H \boldsymbol{\Phi}_r^H \mathbf{a}_i \beta_i \mathbf{a}_i^H \boldsymbol{\Phi}_r \mathbf{G} \boldsymbol{d}_q c_q\left[l \!-\!\tau_i\right]\!+\!\tilde{\mathbf{n}}[l]\right).
\end{equation}

Let $\tilde{\mathbf{y}}_q=\left[\tilde{y}_q\left[1+\tau_q\right], \cdots, \tilde{y}_q\left[L_p+\tau_q\right]\right]^T \in \mathbb{C}^{L_p \times 1}$ denote the signal received within the interval time of all $L_p$ pulses. In the SS scheme, through multiplying $\tilde{\mathbf{y}}_q$ by the original SS sensing signal $\mathbf{c}_q$, DFBS can extract the part corresponding to target $q$ from $\tilde{\mathbf{y}}_q$, that is:

\begin{equation}
	\label{eqn_9}
	\begin{aligned}
		z_q&=\mathbf{c}_q^H \tilde{\mathbf{y}}_q=\sum_{i=1}^Q \alpha_{q, i} \mathbf{c}_q^H \mathbf{c}_i+\mathbf{u}_q^H \sum_{l=1}^{L_p} c_q[l] \tilde{\mathbf{n}}\left[l+\tau_q\right]\\
		&=\alpha_{q,q}\mathbf{c}_q^{H}\mathbf{c}_q+\bar{n}_q=\alpha_{q,q}L_p+\bar{n}_q,
	\end{aligned}
\end{equation}
where $\alpha_{q, i}=\mathbf{u}_q^H \mathbf{G}^H \boldsymbol{\Phi}_r^H \mathbf{a}_i \beta_i \mathbf{a}_i^H \boldsymbol{\Phi}_r \mathbf{G} \boldsymbol{d}_i, \quad \forall i \in \mathcal{Q}, \bar{n}_q=\mathbf{u}_q^H \sum_{l=1}^{L_p} c_q[l] \tilde{\mathbf{n}}\left[l+\tau_q\right]$. Based on \eqref{eqn_9}, we can use the following binary hypothesis test criterion to detect target $q$ \cite{ref36}:

\begin{equation}
	\label{eqn_10}
	z_q=\left\{\begin{array}{l}
		\mathcal{H}_q^0: \bar{n}_q \\
		\mathcal{H}_q^1: \alpha_{q, q} \mathbf{c}_q^H \mathbf{c}_q+\bar{n}_q
	\end{array}\right.,
\end{equation}
where $\mathcal{H}_q^0$ represents there is no target at the $q$-th direction, and $\mathcal{H}_q^1$ represents there exists target at the $q$-th direction. With the above criterion, the optimal detector is $E=\left|z_q\right|^2 \underset{\mathcal{H}_q^1}{\stackrel{\mathcal{H}_q^0}{\lessgtr}} \mu$, where $\mu$ is the threshold satisfying the desired probability of false alarm \cite{ref37,ref38}. Through the aforementioned procedures, it becomes possible to differentiate between the echoes reflected by various targets. It is obvious that the decision's reliability increases as $\left|\alpha_{q, q}\right|$ increases and $\bar{n}_q$ variance decreases. Consequently, it is imperative to maximize the beam pattern gain of sensing beams in each target direction while concurrently constraining interferences. 

\subsection{Communication Model}

Let $\mathbf{h}_k \in \mathbb{C}^{N \times 1}$ represent the channel from the STAR-RIS to the $k$-th user, and the received signal at the time slot $l$ for the $k$-th user can be written as:

\begin{equation}
	\label{eqn_11}
	\begin{aligned}
		&y_k[l]=\mathbf{h}_k^H \boldsymbol{\Phi}_t \mathbf{G} \boldsymbol{x}\left[l-\tau_k\right]+\mathbf{n}_k[l] \\
		& = \mathbf{h}_k^H \boldsymbol{\Phi}_t \mathbf{G}\left(\boldsymbol{\omega}_k s_k\left[l-\tau_k\right]+\sum_{j=1, j \neq k}^{K}\boldsymbol{\omega}_js_j\left[l-\tau_k\right]\right)\\
		&+\mathbf{h}_k^H \boldsymbol{\Phi}_t \mathbf{G} \sum_{q=1}^Q \boldsymbol{d}_q c_q\left[l-\tau_k\right]+\mathbf{n}_k[l].
	\end{aligned}
\end{equation}

Similar to sensing, let $\boldsymbol{y}_k=\left[y_k\left[1+\tau_k\right], \cdots, y_k\left[L_p+\tau_k\right]\right]^T \in \mathbb{C}^{L_p \times 1}$ denote the signal received by user $k$ within the interval time of all $L_p$ pulses, and $\tau_k$ is the propagation delay. Let $\tilde{\mathbf{D}}=\left[\boldsymbol{d}_1, \ldots, \boldsymbol{d}_Q\right] \in \mathbb{C}^{M \times Q}$ and $\tilde{\boldsymbol{c}}[l]=\left[c_1[l], \ldots, c_Q[l]\right]^T \in \mathbb{C}^{Q \times 1}$, thus we have $\tilde{\boldsymbol{c}}[l] \tilde{\boldsymbol{c}}^H[l]=\mathbf{I}_Q$ and \eqref{eqn_11} can be rewritten as:

 \begin{equation}
 	\label{eqn_12}
 	\begin{aligned}
 		\boldsymbol{y}_k[l]& = \underbrace{\hat{\mathbf{h}}_k \boldsymbol{\omega}_k s_k\left[l-\tau_k\right]}_{\text {desired signal }} +\underbrace{\hat{\mathbf{h}}_k\sum_{j=1, j \neq k}^K \boldsymbol{\omega}_j s_j\left[l-\tau_k\right]}_{\text {inter-user interference }} \\
 		& +\underbrace{\hat{\mathbf{h}}_k \tilde{\mathbf{D}} \tilde{\boldsymbol{c}}\left[l-\tau_k\right]}_{\text {sensing interference }}+\mathbf{n}_k[l],
 	\end{aligned}
 \end{equation}
where $\hat{\mathbf{h}}_k=\mathbf{h}_k^H \boldsymbol{\Phi}_t \mathbf{G} \in \mathbb{C}^{1 \times M}$ is the equivalent channel from the DFBS to the $k$-th user at the transmission space, $n_k[l] \sim \mathcal{C N}\left(0, \sigma_k^2\right)$ denotes the complex AWGN with zero mean and variances $\sigma_k^2$ of the $k$-th user. Assume that $\mathbf{D}_q=\boldsymbol{d}_q \boldsymbol{d}_q^H \in \mathbb{C}^{M \times M}$, where $\operatorname{Rank}\left(\mathbf{D}_q\right)=1 \quad$ and $\quad \mathbf{D}_q \succeq 0, \forall q \in \mathcal{Q}$, thus we have $\quad \tilde{\mathbf{D}} \mathbf{I}_Q \tilde{\mathbf{D}}^H=\sum_{q=1}^{\mathcal{Q}} \mathbf{D}_q$ and the corresponding SINR for the $k$-th user can be calculated as:
 
 \begin{equation}
	\label{eqn_13}
	\begin{aligned}
	\operatorname{SINR}_k[l] & =\frac{\left|\hat{\mathbf{h}}_k \boldsymbol{\omega}_k\right|^2}{\sum_{j=1, j \neq k}^K\left|\hat{\mathbf{h}}_k \boldsymbol{\omega}_j\right|^2+\left|\hat{\mathbf{h}}_k \tilde{\mathbf{D}} \tilde{c}\left[l-\tau_k\right]\right|^2+\sigma_k^2} \\
	& =\frac{\left|\hat{\mathbf{h}}_k \boldsymbol{\omega}_k\right|^2}{\sum_{j=1, j \neq k}^K\left|\hat{\mathbf{h}}_k \boldsymbol{\omega}_j\right|^2+\hat{\mathbf{h}}_k \sum_{q=1}^Q \mathbf{D}_q \hat{\mathbf{h}}_k^H+\sigma_k^2}.
	\end{aligned}
\end{equation}

\subsection{Problem Formation}

By adopting SS scheme, it becomes possible to carry out all sensing tasks simultaneously. However, from \eqref{eqn_3} and \eqref{eqn_13} one can find that the transmit beamforming at DFBS and TCs \& RCs at STAR-RIS are both utilized for sensing and communication, which necessitates the efficient balance between the conflicting metrics of sensing and communication. Hence, in this paper, we aim to jointly design the DFBS transmit beamforming $\boldsymbol{\omega}_k, \forall k \in \mathcal{K} $ and $\boldsymbol{d}_q, \forall q \in \mathcal{Q}$, the STAR-RIS reflection beamforming matric $\boldsymbol{\Phi}_r$ and transmission beamforming matric $\boldsymbol{\Phi}_{t}$ to maximize the minimum sensing beam pattern gain under the interference constraint $\eta$, the worst communication SINR requirement $\gamma$, the maximum power budget $P^{\max }$ at the DFBS and the STAR-RIS hardware constraints. The optimization problem thus can be formulated as a max-min fairness problem as follows:

\begin{subequations} \label{eqn_14}
	\begin{align}
		\mathcal P_{0}:
		&\max\limits_{\{\boldsymbol{w}_k\},\left\{\boldsymbol{d}_q\right\}, \mathbf{\Phi}_r, \mathbf{\Phi}_t} \quad \min _{q \in \mathcal{Q}}  \mathcal{P}_q \\
		&\text{s.t.} \hspace{0.3em} \mathrm{C}_{1}: f_{\text {inter}} \leq \eta, \forall q \in \mathcal{Q},\\
		&\hspace{1.5em} \mathrm{C}_{2}: \operatorname{SINR}_k \geq \gamma, \forall k \in \mathcal{K}, \\
		&\hspace{1.5em} \mathrm{C}_{3}: \sum_{k=1}^K\left\|\boldsymbol{\omega}_k\right\|^2+\sum_{q=1}^Q\left\|\boldsymbol{d}_q\right\|^2 \leq P^{\max }, \\
		&\hspace{1.5em} \mathrm{C}_{4}: \beta_{r, n}^2+\beta_{t, n}^2=1, \forall n \in \mathcal{N}, \\
		&\hspace{1.5em} \mathrm{C}_{5}: \beta_{i, n} \in[0,1], \phi_{i, n} \in[0,2 \pi), i \in\{r, t\}, \\
		&\hspace{1.5em} \mathrm{C}_{6}: \boldsymbol{D}_q \succeq \mathbf{0},\forall q \in \mathcal{Q}.
	\end{align}
\end{subequations}

Let $\boldsymbol{\varphi}_i=\left[\beta_{i, 1} e^{j \phi_{i, 1}}, \beta_{i, 2} e^{j \phi_{i, 2}}, \ldots, \beta_{i, N} e^{j \phi_{i, N}}\right]^H, \forall i \in\{t, r\}$, considering the fact that $\mathbf{a}_q^H \boldsymbol{\Phi}_r \mathbf{G}=\boldsymbol{\varphi}_r^H \operatorname{diag}\left(\mathbf{a}_q^H\right) \mathbf{G}=\boldsymbol{\varphi}_r^H \mathbf{A}_q \in \mathbb{C}^{\mathrm{1} \times M}$, where $\mathbf{A}_q=\operatorname{diag}\left(\mathbf{a}_q^H\right) \mathbf{G} \in \mathbb{C}^{N \times M}$, the sensing beam pattern gain \eqref{eqn_3} can be rewritten as $\mathcal{P}_q=\mathbf{a}_q^H \boldsymbol{\Phi} \mathbf{G}_q \boldsymbol{d}_q \boldsymbol{d}_q^H \mathbf{G}^H \boldsymbol{\Phi}_r^H \mathbf{a}_q=\boldsymbol{\varphi}_r^H \mathbf{A}_q \mathbf{D}_q \mathbf{A}_q^H \boldsymbol{\varphi}_r$. For further simplification, we define $\mathbf{W}_k=\boldsymbol{w}_k \boldsymbol{w}_k^H \in \mathbb{C}^{M \times M}$, where $\mathbf{W}_k \succeq \mathbf{0}$ and $\operatorname{Rank}\left(\mathbf{W}_k\right)=1$. Then, $\mathcal P_{0}$ can be transformed into the following one:

\begin{subequations} \label{eqn_15}
	\begin{align}
		\mathcal P_{1}:
		&\max\limits_{\left\{\mathbf{W}_k\right\},\left\{\mathbf{D}_q\right\}, \boldsymbol{\varphi}_r, \boldsymbol{\varphi}_t} \quad \min _{q \in \mathcal{Q}}  \mathcal{P}_q \\
		&\hspace{0.5em} \text{s.t.} \hspace{0.3em} \mathrm{C}_{1}, \mathrm{C}_{2},\\
		&\hspace{2em} \mathrm{C}_{3}: \sum_{k=1}^K \operatorname{Tr}\left(\mathbf{W}_{\mathrm{k}}\right)+\sum_{q=1}^Q \operatorname{Tr}\left(\mathbf{D}_q\right) \leq P^{\max }, \\
		&\hspace{2em} \mathrm{C}_{4},\mathrm{C}_{5},\\
		&\hspace{2em} \mathrm{C}_{6}: \boldsymbol{D}_q \succeq \mathbf{0}, \operatorname{Rank}\left(\mathbf{D}_q\right)=1, \forall q \in \mathcal{Q}, \\
		&\hspace{2em} \mathrm{C}_{7}: \mathbf{W}_{\mathrm{k}} \succeq \mathbf{0}, \operatorname{Rank}\left(\mathbf{W}_{\mathrm{k}}\right)=1,\forall k \in \mathcal{K}.
	\end{align}
\end{subequations}

It is clear that the main difficulty for solving $\mathcal P_{1}$ is the non-convexity induced by the coupled variables in constraints, i.e., the coupling relationship between DFBS transmit beamforming and STAR-RIS TCs \& RCs and STAR-RIS unit-modulus constraint. To tackle this issue, we propose an efficient iterative algorithm in the next section.

\section{Proposed Joint Beamforming Optimization Scheme}

In this section, an $\mathrm{AO}$ scheme is proposed to solve $\mathcal P_{1}$ , namely optimizing $\left(\left\{\mathbf{W}_k\right\},\left\{\mathbf{D}_q\right\}\right)$ and $\left(\boldsymbol{\varphi}_r, \boldsymbol{\varphi}_t\right)$ alternatively until convergence. Firstly, in order to simplify the formulated max-min problem, an auxiliary variable $R$ is introduced as the minimum sensing beam pattern gain, and $\mathcal P_{1}$ can be reformulated as the subsequent equivalent one:

\begin{subequations} \label{eqn_16}
\begin{align}
	\mathcal P_{2}:
	&\max\limits_{\left\{\mathbf{W}_k\right\},\left\{\mathbf{D}_q\right\}, \boldsymbol{\varphi}_r, \boldsymbol{\varphi}_t,R} \quad R \\
	&\hspace{0.5em} \text{s.t.} \hspace{0.3em} \mathrm{C}_{0}:\boldsymbol{\varphi}_r^H \mathbf{A}_q \mathbf{D}_q \mathbf{A}_q^H \boldsymbol{\varphi}_r \geq R, \forall q \in \mathcal{Q},\\
    &\hspace{2em} \mathrm{C}_{1}, \mathrm{C}_{2}, \mathrm{C}_{3}, \mathrm{C}_{4}, \mathrm{C}_{5}, \mathrm{C}_{6}, \mathrm{C}_{7}.
\end{align}
\end{subequations}

\begin{figure*}[ht] %hb代表放在文章底部，%ht为放在文章顶部 
	\centering
	\begin{equation}
		\label{eqn_17}
		\begin{aligned}
			f_{\text {inter }}&=\mathbb{E}\left(\left|\mathbf{a}_q^H \boldsymbol{\Phi}_r \mathbf{G} \sum_{k=1}^K \boldsymbol{w}_k s_k\right|^2\right)+\sum_{q^{\prime} \neq q} \mathbb{E}\left(\left|\mathbf{a}_{q^{\prime}}^H \boldsymbol{\Phi}_r \mathbf{G}\boldsymbol{d}_q c_q\right|^2\right) \\
			& =\boldsymbol{\varphi}_{r}^H \mathbf{A}_q \tilde{\mathbf{W}} \tilde{\boldsymbol{s}} \tilde{\boldsymbol{s}}^H \tilde{\mathbf{W}}^H \mathbf{A}_q \boldsymbol{\varphi}_{r}+\sum_{q^{\prime} \neq q}^Q \mathbb{E}\left(\left|\boldsymbol{\varphi}_{r}^H \mathbf{A}_{q^{\prime}}\boldsymbol{d}_q c_q\right|^2\right) \\
			& =\boldsymbol{\varphi}_r^H \mathbf{A}_q \sum_{k=1}^K \mathbf{W}_{\mathrm{k}} \mathbf{A}_q^H \boldsymbol{\varphi}_{r}+\sum_{q^{\prime} \neq q}^Q \boldsymbol{\varphi}_{r}^H \mathbf{A}_{q^{\prime}}\boldsymbol{d}_q \boldsymbol{d}_q^H \mathbf{A}_{q^{\prime}}^H \boldsymbol{\varphi}_{r} \\
			& =\boldsymbol{\varphi}_r^H \mathbf{A}_q \sum_{k=1}^K \mathbf{W}_{\mathrm{k}} \mathbf{A}_q^H \boldsymbol{\varphi}_{r}+\sum_{q^{\prime} \neq q}^Q \boldsymbol{\varphi}_{r}^H \mathbf{A}_{q^{\prime}}\mathbf{D}_q\mathbf{A}_{q^{\prime}}^H \boldsymbol{\varphi}_{r} \leq \eta, \forall q \in \mathcal{Q}.
		\end{aligned}
	\end{equation}
\end{figure*}

\begin{figure*}[ht] %hb代表放在文章底部，%ht为放在文章顶部 
	\centering
	\begin{equation}
		\label{eqn_18}
		\begin{aligned}
			\operatorname{SINR}_k &=\frac{\left|\hat{\mathbf{h}}_k \boldsymbol{w}_k\right|^2}{\sum_{j=1, j \neq k}^K\left|\hat{\mathbf{h}}_k \boldsymbol{w}_j\right|^2+\hat{\mathbf{h}}_k \sum_{q=1}^Q \mathbf{D}_q \hat{\mathbf{h}}_k^H+\sigma_k^2} =\frac{\boldsymbol{w}_k^H \hat{\mathbf{h}}_k^H \hat{\mathbf{h}}_k \boldsymbol{w}_k}{\sum_{j=1}^K\left|\hat{\mathbf{h}}_k \boldsymbol{w}_j\right|^2-\left|\hat{\mathbf{h}}_k \boldsymbol{w}_k\right|^2+\hat{\mathbf{h}}_k \sum_{q=1}^Q \mathbf{D}_q \hat{\mathbf{h}}_k^H+\sigma_k^2} \\
			& =\frac{\operatorname{Tr}\left(\hat{\mathbf{h}}_k^H \hat{\mathbf{h}}_k \mathbf{W}_k\right)}{\operatorname{Tr}\left(\hat{\mathbf{h}}_k^H \hat{\mathbf{h}}_k \left(\sum_{j=1}^K \mathbf{W}_j+\sum_{q=1}^Q \mathbf{D}_q\right)\right)-\operatorname{Tr}\left(\hat{\mathbf{h}}_k^H \hat{\mathbf{h}}_k \mathbf{W}_k\right)+\sigma_k^2} \geq \gamma.
		\end{aligned}
	\end{equation}
\end{figure*}

\begin{figure*}[ht] %hb代表放在文章底部，%ht为放在文章顶部 
	\centering
	\begin{equation}
		\label{eqn_19}
		\left(1+\gamma^{-1}\right) \operatorname{Tr}\left(\hat{\mathbf{h}}_k^H \hat{\mathbf{h}}_k \mathbf{W}_k\right) \geq \operatorname{Tr}\left(\hat{\mathbf{h}}_k^H \hat{\mathbf{h}}_k\left(\sum_{j=1}^K \mathbf{W}_j+\sum_{q=1}^Q \mathbf{D}_q\right)\right)+\sigma_k^2, \forall k \in \mathcal{K},
	\end{equation}
\end{figure*}
To facilitate the subsequent analysis, we rewrite some constraints of the above optimization problem. For the interference constraint $\mathrm{C}_{1}$, we introduce two auxiliary variables $\tilde{\mathbf{W}}=\left[\boldsymbol{w}_1, \ldots, \boldsymbol{w}_K\right] \in \mathbb{C}^{M \times K}$ and $\tilde{\boldsymbol{s}}=\left[s_1, \ldots, s_K\right]^T \in \mathbb{C}^{K \times 1}$, then there exists $\sum_{k=1}^K \boldsymbol{w}_k s_k=\tilde{\mathbf{W}} \tilde{\boldsymbol{s}}, \quad \tilde{\boldsymbol{s}} \tilde{\boldsymbol{s}}^H \stackrel{(a)}{=} \mathbf{I}_K$ and $\tilde{\mathbf{W}} \tilde{\boldsymbol{s}}^H \tilde{\mathbf{W}}^H \stackrel{(a)}{=} \tilde{\mathbf{W}} \mathbf{I}_K \tilde{\mathbf{W}}^H=\sum_{k=1}^K \mathbf{W}_{k}$, where (a) holds since the previous assumptions $s_k[t] \sim \mathcal{C N}(0,1)$ and $\mathbb{E}\left\{s_k[t] s_{k^{\prime}}^H[t]\right\}=0$, $k \neq k^{\prime}, \forall k, k^{\prime} \in \mathcal{K}$. Thus, the constraint $\mathrm{C}_{1}$ can be written as \eqref{eqn_17} at the top of this page.

For $\mathrm{C}_{2}$, we can calculate it as \eqref{eqn_18} at the top of this page.

By mathematical transformations, $\mathrm{C}_{2}$ can be rewritten as \eqref{eqn_19}, which is presented at the top of the next page.

\subsection{Fix $\left(\boldsymbol{\varphi}_r, \boldsymbol{\varphi}_t\right)$ and Slove $\left(\left\{\mathbf{W}_k\right\},\left\{\mathbf{D}_q\right\}\right)$}

For fixed $\left(\boldsymbol{\varphi}_r, \boldsymbol{\varphi}_t\right)$, the optimization of DFBS transmit beamforming $\left(\left\{\mathbf{W}_k\right\},\left\{\mathbf{D}_q\right\}\right)$ can be modeled as: 

\begin{subequations} \label{eqn_20}
	\begin{align}
		\mathcal P_{3}:
		&\max\limits_{\left\{\mathbf{W}_k\right\},\left\{\mathbf{D}_q\right\},R} \quad R \\
		&\hspace{0.5em} \text{s.t.} \hspace{0.3em} \mathrm{C}_{0},\\
		&\hspace{2em} \mathrm{C}_{1}: (17),\\
		&\hspace{2em} \mathrm{C}_{2}: (19),\\
		&\hspace{2em} \mathrm{C}_{3},\mathrm{C}_{6},\mathrm{C}_{7}.
	\end{align}
\end{subequations}

Obviously, $\mathcal P_{3}$ remains non-convex because of the rank-one constraints. Wherease, $\mathcal P_{3}$ can be relaxed to $\mathcal P_{3.1}$ through the removal of the rank-one constraints:

\begin{subequations} \label{eqn_21}
	\begin{align}
		\mathcal P_{3.1}:
		&\max\limits_{\left\{\mathbf{W}_k\right\},\left\{\mathbf{D}_q\right\},R} \quad R \\
		&\hspace{0.5em} \text{s.t.} \hspace{0.3em} \mathrm{C}_{0},\\
		&\hspace{2em} \mathrm{C}_{1}: (17),\\
		&\hspace{2em} \mathrm{C}_{2}: (19),\\
		&\hspace{2em} \mathrm{C}_{3},\\
		&\hspace{2em} \mathrm{C}_{6}: \mathbf{D}_{\mathrm{q}} \succeq \mathbf{0},\forall q \in \mathcal{Q},\\
	    &\hspace{2em} \mathrm{C}_{7}: \mathbf{W}_{\mathrm{k}} \succeq \mathbf{0}, \forall k \in \mathcal{K}.
	\end{align}
\end{subequations}

It is evident that $\mathcal P_{3.1}$ pertains to a convex QSDP problem, and its optimal solution ( $\left\{\mathbf{W}_k^{\text {opt }}\right\},\left\{\mathbf{D}_q^{\text {opt }}\right\}$ ) can be directly obtained by SDP method. However, after obtaining the optimal solution, we need to check whether they satisfy the rank-one constraint and convert the obtained $\left\{\mathbf{W}_k^{\text {opt }}\right\}$ and $\left\{\mathbf{D}_q^{\text {opt }}\right\}$ into feasible $\left\{\boldsymbol{w}_k^{\text {opt }}\right\}$ and $\left\{\boldsymbol{d}_q^{\text {opt }}\right\}$. Fortunately, according to [20, Theorem 1], [28, Theorem 1] and [35, Proposition 1], the relaxed SDP problm $\mathcal P_{3.1}$ is considered equivalent to $\mathcal P_{3}$ as the the optimality of the solution remains not compromised.

\subsection{Fix $\left(\left\{\mathbf{W}_k\right\},\left\{\mathbf{D}_q\right\}\right)$ and Slove $\left(\boldsymbol{\varphi}_r, \boldsymbol{\varphi}_t\right)$}

For fixed $\left(\left\{\mathbf{W}_k\right\},\left\{\mathbf{D}_q\right\}\right)$, the optimization problem with respect to $\left(\boldsymbol{\varphi}_r, \boldsymbol{\varphi}_t\right)$ can be modeled as:

\begin{subequations} \label{eqn_22}
	\begin{align}
		\mathcal P_{4}:
		&\max\limits_{\boldsymbol{\varphi}_r,\boldsymbol{\varphi}_t,R} \quad R \\
		&\hspace{0.5em} \text{s.t.} \hspace{0.3em} \mathrm{C}_{0},\\
		&\hspace{2em} \mathrm{C}_{1}: (17),\\
		&\hspace{2em} \mathrm{C}_{2}: (19),\\
		&\hspace{2em} \mathrm{C}_{4},\mathrm{C}_{5}.
	\end{align}
\end{subequations}

For analytical tractability, let $\overline{\boldsymbol{\varphi}}_i=\left[\begin{array}{c}\boldsymbol{\varphi}_i \\ u\end{array}\right], \forall i \in\{r, t\}$. Through the matrix transformation, we have $\hat{\mathbf{h}}_k=\mathbf{h}_k^H \boldsymbol{\Phi} \mathbf{G}=\operatorname{diag}\left(\mathbf{h}_k^H\right) \mathbf{G} \boldsymbol{\varphi}_t=\left[\operatorname{diag}\left(\mathbf{h}_k^H\right) \mathbf{G}, \mathbf{0}\right] \overline{\boldsymbol{\varphi}}_t$. By giving the definition \eqref{eqn_23} that is presented at the top of the next page and  $\boldsymbol{\Psi}_i=\overline{\boldsymbol{\varphi}}_i \overline{\boldsymbol{\varphi}}_i^H \in \mathbb{C}^{(N+1) \times(N+1)}$, where $\boldsymbol{\Psi}_i \succeq \mathbf{0}$ and $\operatorname{Rank}\left(\boldsymbol{\Psi}_i\right)=1, \forall i \in\{r, t\}$, we can reformulate $\mathcal P_{4}$ into the explicit and compact forms with respect to $\left(\boldsymbol{\Psi}_r,\boldsymbol{\Psi}_t\right)$ :

\begin{figure*}[ht] %hb代表放在文章底部，%ht为放在文章顶部
	\centering 
	\begin{subequations} \label{eqn_23}
		\begin{align}
			\vspace{0.6em}
			\boldsymbol{B}_q=\left[\begin{array}{cc}
				\mathbf{A}_q \mathbf{D}_q \mathbf{A}_q^H & \mathbf{0} \\
				\mathbf{0} &\mathbf{0}
			\end{array}\right] \in \mathbb{C}^{(N+1) \times(N+1)},\forall q \in \mathcal{Q}, \\
			\vspace{0.6em}
		\boldsymbol{C}_q=\left[\begin{array}{cc}
			\mathbf{A}_q \sum_{k=1}^K \mathbf{W}_{\mathrm{k}} \mathbf{A}_q^H+\sum_{q^{\prime} \neq q}^Q\mathbf{A}_{q^{\prime}}\mathbf{D}_q \mathbf{A}_{q^{\prime}}^H & \mathbf{0} \\
			\mathbf{0} & \mathbf{0}
		\end{array}\right] \in \mathbb{C}^{(N+1) \times(N+1)},\forall q \in \mathcal{Q},\\
			\vspace{0.6em}
			\boldsymbol{E}_k=\left[\begin{array}{cc}
				\operatorname{diag}\left(\mathbf{h}_k^H\right) \mathbf{G} \mathbf{W}_{\mathrm{k}} \mathbf{G}^H \operatorname{diag}^H\left(\mathbf{h}_k^H\right) & \mathbf{0} \\
				\mathbf{0} & \mathbf{0}
			\end{array}\right] \in \mathbb{C}^{(N+1) \times(N+1)},\forall k \in \mathcal{K},\\
			\vspace{0.6em}
			\boldsymbol{F}_k=\left[\begin{array}{cc}
				\operatorname{diag}\left(\mathbf{h}_k^H\right) \mathbf{G}\left(\sum_{k=1}^K \mathbf{W}_{\mathrm{k}}+\sum_{q=1}^Q \mathbf{D}_q\right) \mathbf{G}^H \operatorname{diag}^H\left(\mathbf{h}_k^H\right) & \mathbf{0} \\
				\mathbf{0} & \mathbf{0}
			\end{array}\right] \in \mathbb{C}^{(N+1) \times(N+1)},\forall k \in \mathcal{K}, 
		\end{align}
	\end{subequations}
\end{figure*}

\begin{subequations} \label{eqn_24}
	\begin{align}
		&\mathcal P_{5}:
		\max\limits_{\boldsymbol{\Psi}_{r}, \boldsymbol{\Psi}_{t}, R} \quad R \\
		& \text{s.t.} \mathrm{C}_{0}:\operatorname{Tr}\left(\boldsymbol{B}_q \boldsymbol{\Psi}_r\right) \geq R, \forall q \in \mathcal{Q}, \\
		&\hspace{1.2em} \mathrm{C}_{1}: \operatorname{Tr}\left(\boldsymbol{C}_q \boldsymbol{\Psi}_r\right) \leq \eta, \forall q \in \mathcal{Q}, \\
		&\hspace{1.2em} \mathrm{C}_{2}: \left(1\!+\!\gamma_k^{-1}\right) \operatorname{Tr}\left(\boldsymbol{E}_k \boldsymbol{\Psi}_t\right) \geq \operatorname{Tr}\left(\boldsymbol{F}_k \boldsymbol{\Psi}_t\right)\!+\!\sigma_k^2, \forall k \in \mathcal{K},\\
		&\hspace{1.2em} \mathrm{C}_{4}:{\left[\boldsymbol{\Psi}_r\right]_{n, n}\!+\!\left[\boldsymbol{\Psi}_t\right]_{n, n}\!=\!1,\left[\boldsymbol{\Psi}_{r/t}\right]_{n, n} \geq 0, \forall n \in \mathcal{N}}, \\
		&\hspace{1.2em} \mathrm{C}_{8}:\boldsymbol{\Psi}_r \succeq \mathbf{0}, \operatorname{Rank}\left(\boldsymbol{\Psi}_r\right)=1,\\
		&\hspace{1.2em} \mathrm{C}_{9}:\boldsymbol{\Psi}_t \succeq \mathbf{0}, \operatorname{Rank}\left(\boldsymbol{\Psi}_t\right)=1.
	\end{align}
\end{subequations}

Obviously, rank-one constraints (24f) and (24g) result in $\mathcal P_{5}$ being non-convex, we utilize the SDR method to relax it, and $\mathcal P_{5}$ can be reduced to:

\begin{subequations} \label{eqn_25}
	\begin{align}
		\mathcal P_{5.1}:
		&\max\limits_{\boldsymbol{\Psi}_{r}, \boldsymbol{\Psi}_{t}, R} \quad R \\
	    &\hspace{0.5em} \text{s.t.} \hspace{0.3em} \mathrm{C}_{0},\mathrm{C}_{1},\mathrm{C}_{2},\mathrm{C}_{4} \\
		&\hspace{2em} \mathrm{C}_{8}:\boldsymbol{\Psi}_r \succeq \mathbf{0},\\
		&\hspace{2em} \mathrm{C}_{9}:\boldsymbol{\Psi}_t \succeq \mathbf{0}.
	\end{align}
\end{subequations}

Similar to the solutions of $\mathcal P_{3.1}$ for the rank-one constraints, $\mathcal P_{5.1}$ can also be solved by adopting SDP method. However, when iteratively solving $\mathcal P_{3.1}$ and $\mathcal P_{5.1}$ until they converge and obtain the optimized solutions, we still need to check whether they satisfy the rank-one constraint and convert the obtained $\boldsymbol{\Psi_i}^{\text {opt}}$ into feasible $\boldsymbol{\varphi}_i^{o p t}, \forall i \in\{r, t\}$. Specifically, Gaussian randomization or eigenvalue decomposition methods can be used to approximate the relaxed rank-one constraint solution and obtain the ultimate $\boldsymbol{\Phi}_r$ and $\boldsymbol{\Phi}_t$ \cite{ref39,ref40,ref41}.

\begin{algorithm}[H]
	\caption{Joint beamforming design for STAR-RIS-enabled ISAC system}\label{alg:alg1}
	\begin{algorithmic}
		\STATE 
		\STATE \textbf{Input}: $\mathbf{h}_k,\sigma_k^2,\mathbf{a}_q,\mathbf{G},\sigma_z^2, P^{\max }, \gamma, \eta, \Gamma_{\max }, \delta_{t h}.$  
		\STATE \textbf{Output}: DFBS beamforming $\left\{\mathbf{W}_k\right\}$ and $\left\{\mathbf{D}_q\right\}$, STAR-RIS TCs
		$\boldsymbol{\Phi}_r$ and RCs $\boldsymbol{\Phi}_t$ and optimized minimum beam pattern gain $\mathcal{P}_{\text {gain}}$.
		\STATE 1: Initialize: $\left\{\mathbf{W}_k\right\},\left\{\mathbf{D}_q\right\}, \boldsymbol{\Phi}_{\mathrm{t/r}},i\!=\!1, \delta\!=\!\infty,R\!=\!0$
		\STATE 2: $\textbf{while}$ $i \leq \Gamma_{\max }$ and $\delta \geq \delta_{t h}$  $\textbf{do}$
		\STATE 3: \hspace{0.7cm} $\mathcal{P}_{\text {pre }}=R$;     
		\STATE 4: \hspace{0.7cm} Optimize $\left\{\mathbf{W}_k\right\}$ and $\left\{\mathbf{D}_q\right\}$ by solving  $\mathcal P_{3.1}$;
		\STATE 5: \hspace{0.7cm} Optimize $\boldsymbol{\Psi}_r$ and $\boldsymbol{\Psi}_t$ by solving $\mathcal P_{5.1}$;
		\STATE 6: \hspace{0.7cm} $\mathcal{P}_{\text {gain}}=R$;
		\STATE 7: \hspace{0.7cm} $\delta=\frac{\left|\mathcal{P}_{\text {pre }}-\mathcal{P}_{\text {gain}}\right|}{\mathcal{P}_{\text {gain}}}$;
		\STATE 8: \hspace{0.6cm} $i=i+1$.
		\STATE 9: $\textbf{end while}$      		
	\end{algorithmic}
	\label{alg1}
\end{algorithm}

Summarily, to solve $\mathcal P_{0}$, we first initialize the feasible variables $\left\{\mathbf{W}_k\right\},\left\{\mathbf{D}_q\right\}, \boldsymbol{\varphi}_t$ and $\boldsymbol{\varphi}_r$, and solve $\mathcal P_{3.1}$ to obtain $\left\{\mathbf{W}_k^{\text {opt }}\right\}$ and $\left\{\mathbf{D}_q^{\text {opt }}\right\}$, and then solve $\mathcal P_{5.1}$ to obtain $\boldsymbol{\Psi_r}^{\text {opt}}$ and $\boldsymbol{\Psi_t}^{\text {opt}}$, and then repeat $\mathcal P_{3.1}$ and $\mathcal P_{5.1}$ until the minimum sensing beam patter gain converges to a stable value, which is summarized as Algorithm 1.

\subsection{Analysis of Convergence and Computional Complexity}

The proposed joint beamforming scheme in Algorithm 1 requires iterative solutions for $\mathcal P_{3.1}$ and $\mathcal P_{5.1}$ until convergence. As the rank-one constraints are omitted, both $\mathcal P_{3.1}$ and $\mathcal P_{5.1}$ are convex optimization problems, ensuring the existence of Karush-Kuhn-Tucker (KKT) solutions. Furthermore, the limited transmission power makes $\left\{\mathbf{W}_k\right\}$ and $\left\{\mathbf{D}_q\right\}$ bounded, and the amplitude coupling and phase shift unit module constraints also bound the TCs \& RCs of STAR-RIS, collectively resulting in an upper bound for $\mathcal P_{0}$. Consequently, the beam pattern gain under Algorithm 1 is expected to exhibit monotonically non-decreasing behavior and converge to a local optimal solution. 

\begin{figure}[!t]
	\centering
	\includegraphics[height=2.8in,width=3.4in]{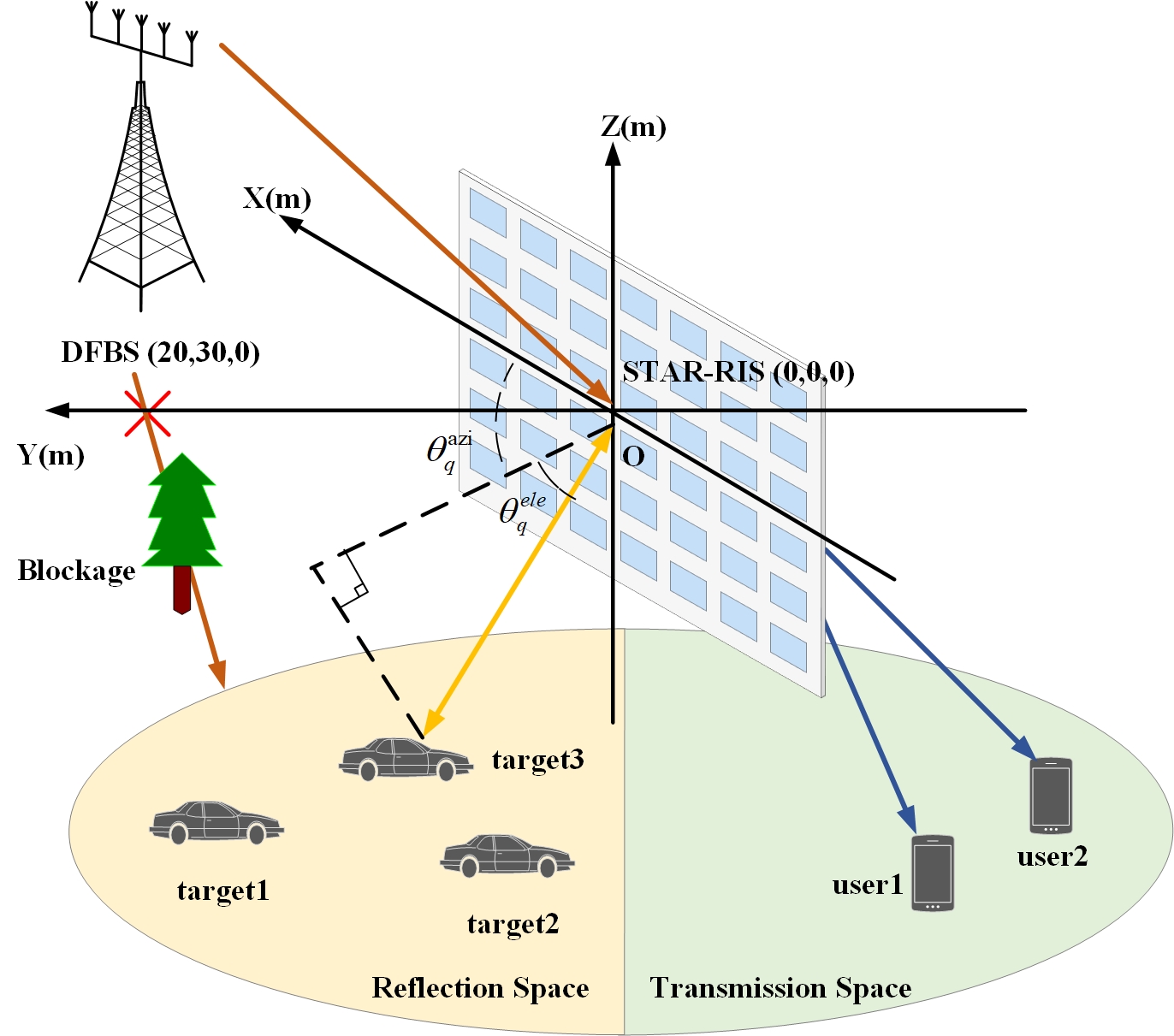}
	\caption{Simulation setup for STAR-RIS-enabled ISAC system.}
	\label{fig_2}
\end{figure}

Next, we provide a concise analysis of the computional complexity. With reference to the analysis in [39], first, we give an iteration precision $\varDelta$, and the number of iterations necessary for Algorithm 1 to achieve convergence can be expressed as $\mathrm{log}(1/\varDelta)$. Thus, in each iteration, given the complexity of finding a solution for the SDR problem, the computational complexity of optimizing DFBS transmit beamforming $\left(\left\{\mathbf{W}_k\right\},\left\{\mathbf{D}_q\right\}\right)$ is $\mathcal{O}\left(\max \{M, K+Q\}^4 \sqrt{M} \right)$ and optimizing STAR-RIS TCs \& RCs $\left(\boldsymbol{\Phi}_r, \boldsymbol{\Phi}_t\right)$ is $\mathcal{O}\left(\max \{N+1, K+Q\}^4 \sqrt{N+1} \right)$. Thus, the overall computational complexity of Algorithm 1 is $\mathcal{O}\left(\mathrm{log}(1/\varDelta)\left(\max \{M, K+Q\}^4 \sqrt{M} +\right.\right.$ $\left.\left.\max \{N+1, K+Q\}^4 \sqrt{N+1} \right)\right)$.

\begin{table}[ht]
	\caption{System Parameters\label{tab:table1}}
	\centering
	\resizebox{0.5\textwidth}{!}{
		\begin{tabular}{|c||c|}
			\hline
			Paramater & Value\\
			\hline
			The number of antennas at DFBS & $M=6$\\
			\hline
			The number of STAR-RIS elements & $N=40$\\
			\hline
			The number of communication users & $k=2$\\
			\hline
			The number of targets & $Q=3$\\
			\hline
			Path loss exponents & $\tau_{BR}=\tau_{RU,k}=1$\\
			\hline
			The path loss at 1 m & $L_0=30$dB\\
			\hline
			Rician factor & $\kappa$=1\\
			\hline
			Transmit power at the DFBS & $P^{\max }=10 \mathrm{dBm}$ \\
			\hline
			The noise power & $\sigma^2=-110 \mathrm{dBm}$ \\
			\hline
			Azimuth and elevation DOAs of the targets & $(120^{\circ},30^{\circ}),(60^{\circ},60^{\circ})$ \\
			&and $(30^{\circ},75^{\circ})$  \\
			\hline
			The interference threshold & $\eta = 10^{-3}$ \\
			\hline
			The communication SINR requirement & $\gamma = 6$dB \\
			\hline
		\end{tabular}
	}
\end{table}

\section{Simulation Results}

In this section, simulation results are conducted to evaluate the performance of the proposed scheme. As depicted in Fig. \ref{fig_2}, in order to accurately depict the locations of DFBS and STAR-RIS, a 3D coordinate system is utilized. Concretely, the STAR-RIS is positioned at the origin $(0, 0 , 0)$ meters$(\mathrm{m})$, while the DFBS is $30 \mathrm{~m}$ away from STAR-RIS, and its plane coordinate is $(20, 30, 0)\mathrm{m}$. The sensing targets are situated within the reflection space and the communication users are randomly distributed between $30\mathrm{m}$ and $50\mathrm{m}$ within the transmission space. We set the number of antennas at DFBS to $M=6$ and the transmit power to $P^{\max }=10 \mathrm{dBm}$. The number of communication users is set as $K=2$. The number of targets is set as $Q=3$, and the azimuth and elevation DOAs of targets are $\left(120^{\circ}, 30^{\circ}\right),\left(60^{\circ}, 60^{\circ}\right)$ and $\left(30^{\circ}, 75^{\circ}\right)$, respectively. For simplicity, we set all noise power as the same, i.e., $\sigma_z^2=\sigma_k^2=-110$ $\mathrm{dBm}, \forall k$. Assuming that the number of STAR-RIS elements is $N=N_x \times N_z=8 \times 5=40$. Without loss of generality, the Rician channel model is adopted for all channels. Hence, $\mathbf{G} \in \mathbb{C}^{N \times M}$ and $\mathbf{h}_k \in \mathbb{C}^{N \times 1}, \forall k$ are computed with $\mathbf{G} =\sqrt{L_0\left(\frac{d_{BR}}{d_0}\right)^{\!-\!\tau_{BR}}}\left(\sqrt{\frac{\kappa}{\kappa\!+\!1}} \mathbf{G}^{\text {LoS}}\!+\!\sqrt{\frac{1}{\kappa\!+\!1}} \mathbf{G}^{NLoS}\right)$ and $\mathbf{h}_k =\sqrt{L_0\left(\frac{d_{RU,k}}{d_0}\right)^{\!-\!\tau_{RU,k}}}\left(\sqrt{\frac{\kappa}{\kappa\!+\!1}} \mathbf{h}_k^{\text {LoS }}\!+\!\sqrt{\frac{1}{\kappa\!+\!1}} \mathbf{h}_k^{NLoS}\right)$, respectively, where $L_0$ represents the path loss of the reference distance $d_0=1 \mathrm{~m}, d_{B R}$ and $d_{RU, k}$, respectively, represent the distance between the DFBS and the STAR-RIS, and the distance between the STAR-RIS and the $k$-th user. $\tau_{BR}$ and $\tau_{RU,k}$ denote the corresponding path loss exponents. $\kappa$ is the Rician factor, in particular, here we set $\kappa=1$. The specific parameters setting can be found in Table I.

\begin{figure*}[!t]
	\centering
	\subfloat[]{\includegraphics[height=2.8in,width=3.2in]{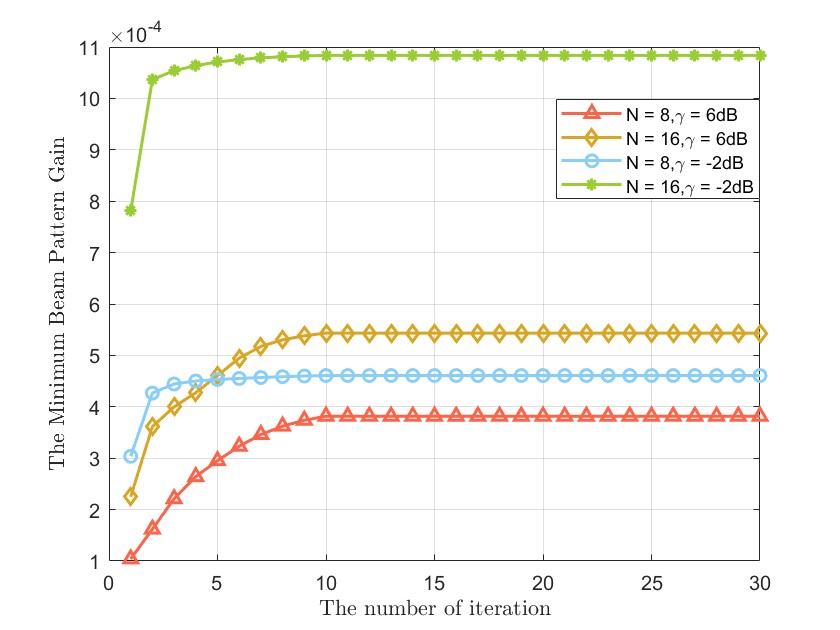}%
		\label{(fig_first_case)}}
	\hfil
	\subfloat[]{\includegraphics[height=2.8in,width=3.2in]{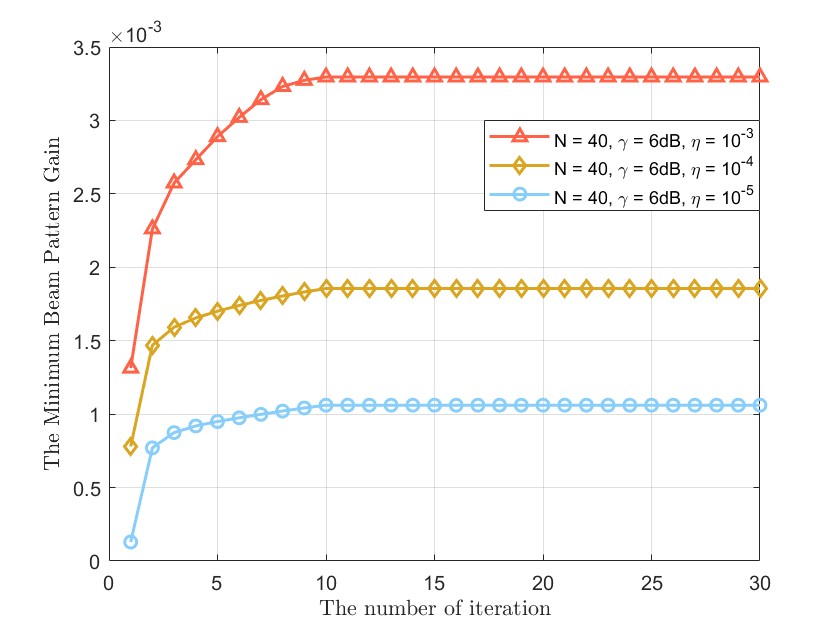}%
		\label{fig_second_case}}
	\caption{Beam pattern gain versus the number of iterations. (a) With various $\gamma$ and $N$. (b) With various $\eta$.}
	\label{fig_3}
\end{figure*}

\begin{figure*}[!t]
	\subfloat[]{\includegraphics[height=2.8in,width=3.2in]{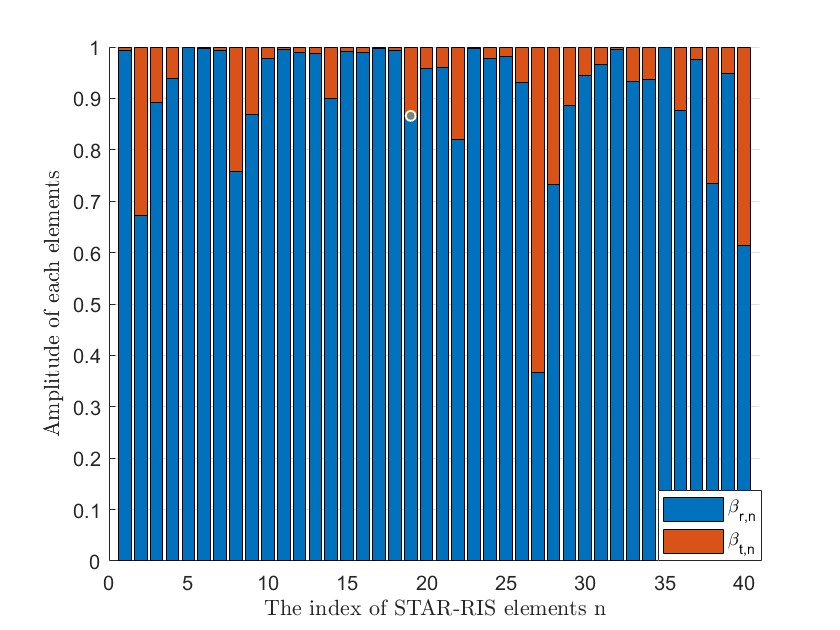}%
		\label{(fig_first_case)}}
	\hfil
	\subfloat[]{\includegraphics[height=2.8in,width=3.2in]{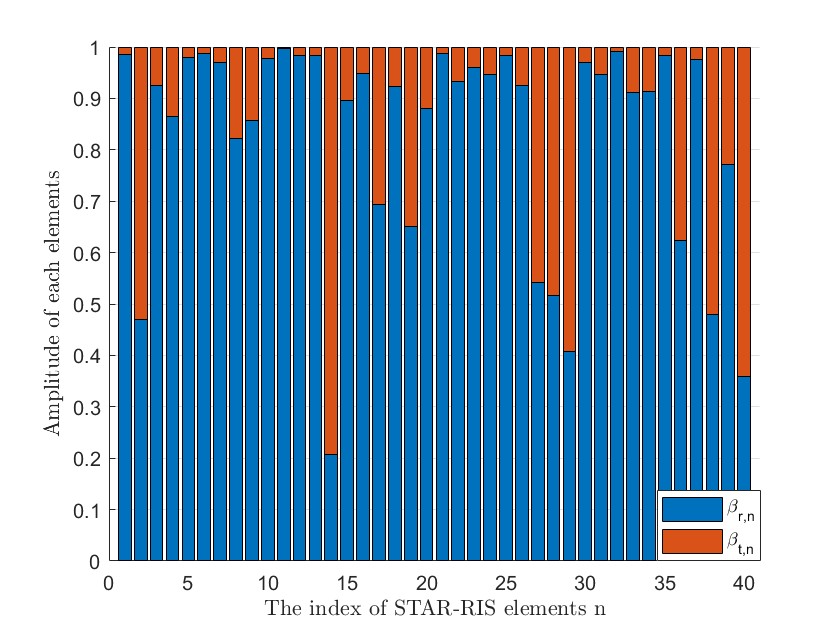}%
		\label{fig_second_case}}
	\caption{The amplitude of each STAR-RIS element under different communication thresholds. (a) $\gamma=10$dB, $N=40$. (b) $\gamma=16$dB, $N=40$.}
	\label{fig_4}
\end{figure*}

\begin{figure}[!t]
	\centering
	\includegraphics[height=2.8in,width=3.2in]{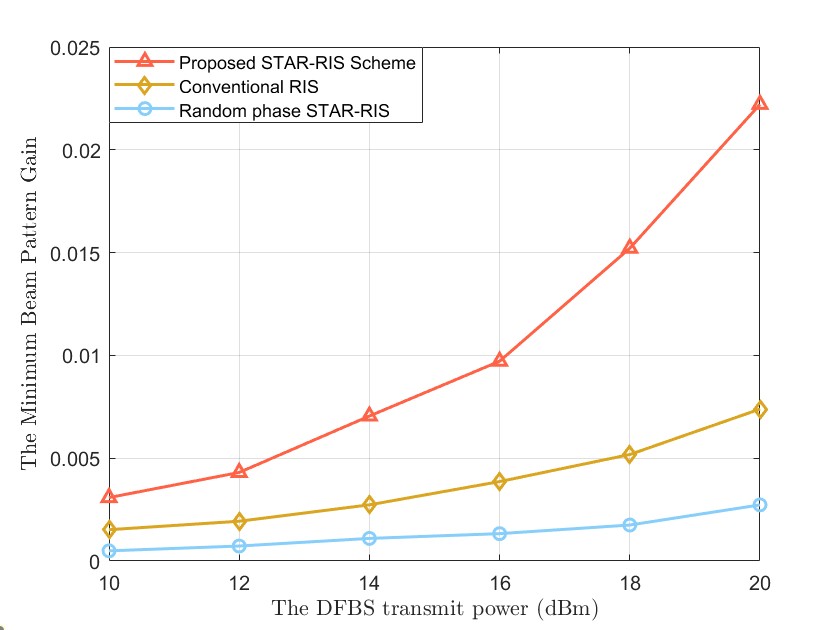}
	\caption{Beam pattern gain versus the DFBS transmit power $P^{\text{max}}$.}
	\label{fig_5}
\end{figure}

\begin{figure}[!t]
	\centering
	\includegraphics[height=2.8in,width=3.2in]{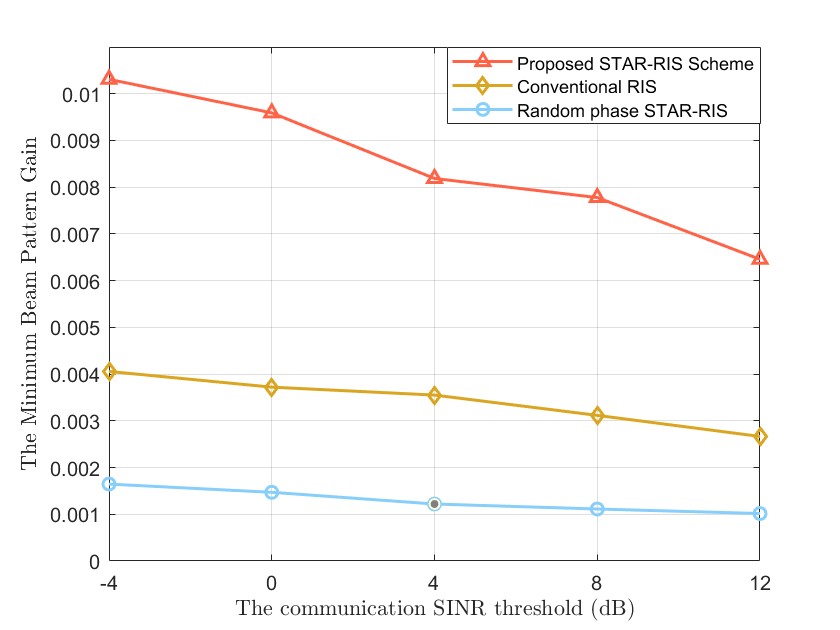}
	\caption{Beam pattern gain versus the communication threshold $\gamma$ ($P^{\text{max}}=16$dBm).}
	\label{fig_6}
\end{figure}

\begin{figure}[!t]
	\centering
	\includegraphics[height=2.8in,width=3.2in]{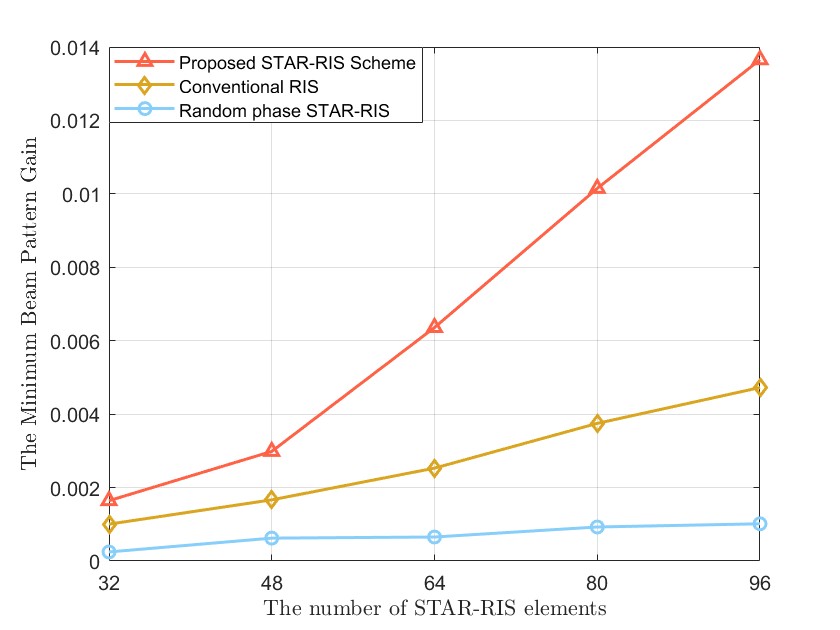}
	\caption{Beam pattern gain versus the number of STAR-RIS elements $N$.}
	\label{fig_7}
\end{figure}

To verify the effectiveness of our proposed scheme, we conduct a comparison with the following baselines:

\begin{itemize}
	\item{\textbf{The random phase shift STAR-RIS-enabled ISAC system scheme (legend ``Random phase STAR-RIS'')}: The initial phase shift of STAR-RIS, i.e., TCs \& RCs are random. Based on this, only the DFBS beamforming optimization is considered.}
	
	\item{\textbf{The conventional RIS-enabled ISAC system scheme (legend ``Conventional RIS'')}: Two conventional RISs, one designed for reflecting-only and the other designed for transmission-only, are both assumed with $N / 2$ elements and deployed adjacent to each other at the same location as the STAR-RIS for the purpose of fair comparisons [20], [28].} 
\end{itemize}

Additionally, for the sake of illustration, the following simulation analysis takes the absolute value of the minimum beam pattern gain as the horizontal axis.

1) \textit{Relationship between the minimum beam pattern gain and the number of iterations}: Fig. \ref{fig_3} shows the convergence of Algorithm 1 with various system parameters. Specifically, Fig. \ref{fig_3}(a) and Fig. \ref{fig_3}(b) examine the convergence of Algorithm 1 under different parameter configurations of the number of STAR-RIS elements $N$, the communication SINR requirement $\gamma$ and the interference constraint $\eta$. We can observe that the minimum beam pattern gain rapidly converges to stabilization within a small number of iterations for any given $\gamma, N$ and $\eta$, which validates the effectiveness of our proposed scheme. In addition, Fig. \ref{fig_3}(b) further demonstrates that as the interference constraint becomes stricter, the minimum beam pattern gain decreases. For a specific number of targets, one can find that the minimum beam pattern gain decreases as the interference constraints are reduced from $10^{-3}$ to $10^{-5}$, owing to the imposition of more stringent interference limitations.

2) \textit{Relationship between the amplitude of each STAR-RIS element and the communication threshold}: Fig. \ref{fig_4} illustrates the amplitude of each STAR-RIS element across various communication thresholds. As depicted in the figure, each STAR-RIS element owns the capability to adapt the incident signal in response to its specific propagation environments, and the total of the squared amplitude of reflection and transmission for each STAR-RIS element equals 1. Additionally, it can be visually observed that as the communication requirement becomes stricter, a greater number of amplitudes at the STAR-RIS is allocated to the transmission space. This can be attributed to the heightened communication demands per user associated with a larger $\gamma$, prompting the STAR-RIS to allocate more energy to the transmission space to improve the signal strength.

3) \textit{Relationship between the minimum beam pattern gain and DFBS transmit power}: Fig. \ref{fig_5} provides the relation between the minimum beam pattern gain and the maximum transmit power $P^{\max }$ across various schemes. We can note that the minimum beam pattern gain exhibits a gradual rise with $P^{\max }$ increasing under all schemes. Obviously, our proposed scheme owns the highest beam pattern gain compared to other schemes, and this suggests that under the requirements of simultaneously meeting the constraints for sensing and communication, the proposed scheme typically requires less $P^{\max }$ than other schemes. This can be explained as follows: firstly, the conventional RISs usually utilize fixed number of elements, limiting the ability to control the number of reflection and transmission elements according to different situations as in the case of STAR-RIS. Therefore, STAR-RIS can achieve a better degree of freedom (DoF) to improve the expected signal strength. Additionally, in contrast to the scheme with random phase shift, the proposed scheme can dynamically optimize TCs and RCs of STAR-RIS to more effectively allocate power for signal propagation.

4) \textit{The minimum beam pattern gain versus communication threshold}: We plot Fig. \ref{fig_6} to show the minimum beam pattern gain versus the communication SINR threshold $\gamma$ across various schemes. It can be found that the minimum beam pattern gain under all schemes decreases as $\gamma$ increases, and the proposed scheme obtains the highest beam pattern gain. This occurs as an attribution to allocate more power for communication at higher SINR requirements, resulting in a reduction of power available for target sensing and consequently leading to a decrease of the minimum beam pattern gain. This observation highlights the tradeoff between communication and sensing within the limited resources, and the realization of high communication performance comes at the cost of diminished sensing performance.

5) \textit{Relationship between the minimum beam pattern gain and the number of STAR-RIS elements}: Fig. \ref{fig_7} plots the minimum beam pattern gain versus the number of STAR-RIS elements. We can find that a larger $N$ can obtain a higher beam pattern gain under all schemes, but the proposed scheme owns the best performance. This is easy to understand, as more STAR-RIS elements can achieve a higher passive beamforming gain, and the proposed scheme benefits from its flexible beamforming DoF. Therefore, when $N$ increases, compared to other schemes $N$ can be selected flexibly to satisfy the sensing requirement more effectively.

\section{Conclusion}
In this paper, under the purpose of maximizing the minimum sensing beam pattern gain, we studied an ISAC system with multiple targets and multiple users assisted by STAR-RIS. Specifically, the SS modulation method is first incorporated into the system to address the challenge of DFBS's inability to differentiate among targets, and an analysis is conducted to explain how DFBS distinguishes different targets. Then, we designed a joint beamforming optimization problem of DFBS and STAR-RIS and formulated it into a max-min problem. By applying OA method, the formulated problem was divided into two QSDP sub-problems, and we solved them through SDR and SDP methods. The obtained simulation results provided validation for the benefits of ISAC system enabled by STAR-RIS, as well as the efficacy of the proposed scheme. In future, we will explore the scenario with imperfect channel state information to enhance the practicality of our research.


\begin{thebibliography}{99}
\bibliographystyle{IEEEtran}

\bibitem{ref1}
F. Liu, Y. Cui, C. Masouros, J. Xu, T. X. Han, Y. C. Eldar, and S. Buzzi, ``Integrated sensing and communications: Toward dual-functional wireless networks for 6G and beyond,'' \textit{IEEE J. Sel. Areas Commun.}, vol. 40, no. 6, pp. 1728--1767, 2022.

\bibitem{ref2}
 F. Liu, C. Masouros, A. P. Petropulu, H. Griffiths, and L. Hanzo, ``Joint radar and communication design: Applications, state-of-the-art, and the road ahead,'' \textit{IEEE Trans. Commun.}, vol. 68, no. 6, pp. 3834--3862, Jun. 2020.
 
 \bibitem{ref3}
 J. A. Zhang, M. L. Rahman, K. Wu, X. Huang, Y. J. Guo, S. Chen, and J. Yuan, ``Enabling joint communication and radar sensing in mobile networks—a survey,'' \textit{IEEE Commun. Surv. Tut.}, vol. 24, no. 1, pp. 306--345, 1st Quart. 2021.
 
 \bibitem{ref4}
 I. Bekkerman and J. Tabrikian, ``Target detection and localization using MIMO radars and sonars,'' \textit{ IEEE Trans. Signal Process.}, vol. 54, no. 10, pp. 3873--3883, Oct. 2006.
 
 \bibitem{ref5}
 R. Liu, M. Liu, H. Luo, Q. Liu, and A. L. Swindlehurst, ``Integrated Sensing and Communication with Reconfigurable Intelligent Surfaces: Opportunities, Applications, and Future Directions,'' \textit{IEEE Wireless Commun.}, vol. 30, no. 1, pp. 50--57, Feb. 2023.
 
 \bibitem{ref6}
 M. D. Renzo, A. Zappone, M. Debbah, M. S. Alouini, S. Tretyakov, ``Smart radio environments empowered by reconfigurable intelligent surfaces: How it works, state of research, and the road ahead,'' \textit{IEEE J. Sel. Areas Commun.}, vol. 38, no. 11, pp. 2450--2525, Jul. 2020.
 
 \bibitem{ref7}
 S. Basharat, S. A. Hassan, H. Pervaiz, A. Mahmood, Z. Ding, and M. Gidlund, ``Reconfigurable intelligent surfaces: Potentials, applications, and challenges for 6G wireless networks,'' \textit{IEEE Wirel. Commun.}, vol. 28, no. 6, pp. 184--191, Jul. 2021.
 
 \bibitem{ref8}
 Q. Wu, S. Zhang, B. Xiong, C. You, and R. Zhang, ``Intelligent reflecting surface aided wireless communications: A tutorial,'' \textit{IEEE Trans. Commun.}, vol. 69, no. 5, pp. 3313--3351, May. 2021.
 
 \bibitem{ref9}
 F. Wang, H. Li, and J. Fang, ``Joint active and passive beamforming for IRS-assisted radar,''\textit{ IEEE Signal Process. Lett.}, vol. 29, pp. 349--353, Dec. 2021.
 
  \bibitem{ref10}
 X. Song, D. Zhao, H. Hua, T. Xiao Han, X. Yang, and J. Xu, ``Joint transmit and reflective beamforming for IRS-assisted integrated sensing and communication,'' in \textit{Proc. of the 2022 IEEE Wireless Commun. Netw. Conf. (WCNC)}, Austin, TX, USA, May. 2022.

\bibitem{ref11}
Z. M. Jiang, M. Rihan, P. Zhang, L. Huang, Q. Deng, J. Zhang, and E. M. Mohamed, ``Intelligent reflecting surface aided dual-function radar and communication system,'' \textit{IEEE Syst. J.}, vol. 16, no. 1, pp. 475--486, Mar. 2022.

\bibitem{ref12}
R. Liu, M. Liu, and A. L. Swindlehurst, ``Joint beamforming and reflection design for RIS-assisted ISAC systems,'' in \textit{Proc. of the 2022 30th European Signal Processing Conf. (EUSIPCO)}, Belgrade, Serbia, Oct. 2022.

\bibitem{ref13}
X. Song, D. Zhao, H. Hua, T. Xiao Han, X. Yang, and J. Xu, ``Joint transmit and reflective beamforming for IRS-assisted integrated sensing and communication,'' in \textit{Proc. of the 2022 IEEE Wireless Commun. Netw. Conf. (WCNC)}, Austin, TX, USA, May. 2022.

\bibitem{ref14}
H. Luo, R. Liu, M. Li, and Q. Liu, ``RIS-aided integrated sensing and communication: joint beamforming and reflection design,'' \textit{ IEEE Trans. Veh. Technol.}, vol. 72, no. 7, pp. 9626--9630, Mar. 2022.

\bibitem{ref15}
Y. Liu et al., ``STAR: Simultaneous transmission and reflection for 360◦ coverage by intelligent surfaces,'' \textit{IEEE Wireless Commun.}, vol. 28, no. 6, pp. 102--109, Dec. 2021.

\bibitem{ref16}
S. Zhang et al., ``Intelligent omni-surfaces: Ubiquitous wireless transmission by reflective-refractive metasurfaces,'' \textit{IEEE Trans. Wireless Commun.}, vol. 21, no. 1, pp. 219--233, Jan. 2022.

\bibitem{ref17}
J. Xu, Y. Liu, X. Mu, R. Schober, and H. V. Poor, ``STAR-RISs: A correlated T$\&$R phase-shift model and practical phase-shift configuration strategies,'' \textit{IEEE J. Sel. Topics Signal Process.}, vol. 16, no. 5, pp. 1097--1111, Aug. 2022.

\bibitem{ref18}
Y. Liu, X. Mu, R. Schober, and H. V. Poor, ``Simultaneously transmitting and reflecting (STAR)-RISs: A coupled phase-shift model,'' in \textit{Proc. of IEEE ICC 2022}, Seoul, Korea, Republic of, May. 2022, pp. 2840--2845.

\bibitem{ref19}
B. O. Zhu et al., ``Dynamic control of electromagnetic wave propagation with the equivalent principle inspired tunable metasurface,'' \textit{Sci. Rep.}, vol. 4, no. 1, May. 2014.

\bibitem{ref20}
X. Mu, Y. Liu, L. Guo, J. Lin, and R. Schober, ``Simultaneously transmitting and reflecting (STAR) RIS aided wireless communications,'' \textit{IEEE Trans. Wireless Commun.}, vol. 21, no. 5, pp. 3083--3098, May. 2022.

\bibitem{ref21}
W. Cai, M. Li, Y. Liu, L. Q. Wu, and Q. Liu, ``Joint beamforming design for intelligent omni surface assisted wireless communication systems,'' \textit{IEEE Trans. Wireless Commun.}, vol. 22, no. 2, pp. 1281--1297, Feb. 2023.

\bibitem{ref22}
Y. Chen, Y. Wang, Z. Wang, and P. Zhang, ``Robust beamforming for active reconfigurable intelligent omni-surface in vehicular communications,'' \textit{IEEE J. Sel. Areas Commun.}, vol. 40, no. 10, pp. 3086--3103, Oct. 2022.

\bibitem{ref23}
Y. Ma, M. Li, Y. Liu, Q. Wu, and Q. Liu, ``Optimization for reflection and transmission dual-functional active RIS-assisted systems,'' \textit{IEEE Trans. Commun.}, vol. 71, no. 9, pp. 5534--5548, Jun. 2023.

\bibitem{ref24}
H. Niu and X. Liang, ``Weighted sum-rate maximization for STAR-RISs-aided networks with coupled phase-shifters,'' \textit{IEEE Systems J.}, vol. 17, no. 1, pp. 1083--1086, Mar. 2023.

\bibitem{ref25}
H. Luo, L. Lv, Q. Wu, Z. Ding, N. A. Dhahir, and J Chen, ``Beamforming design for active IOS aided NOMA networks,'' \textit{IEEE Wireless Commun. Lett.}, vol. 12, no. 2, pp. 282--286, Feb. 2023.

\bibitem{ref26}
Z. Wang, X. Mu, J. Xu, and Y. Liu, ``Simultaneously Transmitting and Reflecting Surface (STARS) for Terahertz Communications,'' \textit{IEEE J. Sel. Topics Signal Process.}, vol. 17, no. 4, pp. 861--877, May. 2023.

\bibitem{ref27}
Z. Wang, X. Mu, and Y. Liu, ``STARS enabled integrated sensing and communications,'' \textit{IEEE Trans. Wireless Commun.}, vol. 22, no. 10, pp. 6750--6765, Feb. 2023.

\bibitem{ref28}
H. Luo, R. Liu, M. Li, and Q. Liu, ``Toward STAR-RIS-Empowered Integrated Sensing and Communications: Joint Active and Passive Beamforming Design,'' \textit{IEEE Trans. Veh. Technol.}, early access, pp. 1--15, Jul. 2023.

\bibitem{ref29}
Z. Liu, X. Li, H. Ji, and H. Zhang, ``Exploiting STAR-RIS for physical layer security in integrated sensing and communication networks,'' in \textit{Proc. of the  IEEE 34th Annual International Symposium on Personal, Indoor and Mobile Radio Commun.}, Jun. 2023, pp. 1--8.

\bibitem{ref30}
W. Chao, C. Wang, Z. Li, D. Ng, G, and K. Wong, ``PHY Security Enhancement Exploiting STAR-RIS for Dual-functional Radar-Communication,'' in \textit{Proc. of 2023 IEEE International Conference on Communications Workshops (ICC Workshops)}, Oct. 2023, pp. 1--5.

\bibitem{ref31}
Z. Zhu, M. Gong, Y. Liu, Z. Chu, P. Xiao, G. Sun, D. Mi, Z. He, and F. Tong, ``DRL-based STAR-RIS-assisted ISAC secure communications,'' in \textit{Proc. of the International Conference on Ubiquitous Commun. 2023 (Ucom 2023)}, Xi'an, China, Jun. 2023, pp. 1--6.

\bibitem{ref32}
A. Aubry, A. De Maio, and M. Rosamilia, ``RIS-aided radar sensing in N-LOS environment,'' in \textit{Proc. of IEEE 8th Int. Workshop Metrology Aerosp. (MetroAeroSpace)},Jun. 2021, pp. 277--282.

\bibitem{ref33}
L. Xu, J. Li, and P. Stoica, ``Radar imaging via adaptive MIMO techniques,'' in \textit{Proc. of EUSIPCO}, Sep. 2006, pp. 1--5.

\bibitem{ref34}
P. Stoica, J. Li, and Y. Xie, ``On probing signal design for MIMO radar,'' \textit{IEEE Trans. Signal Process.}, vol. 55, no. 8, pp. 4151--4161, Aug. 2007.

\bibitem{ref35}
K. Meng, Q. Wu, R. Schober, and W. Chen, ``Intelligent Reflecting Surface Enabled Multi-Target Sensing,'' \textit{IEEE Trans. Commun.}, vol. 70, no. 12, pp. 8313--8330, Dec. 2022.

\bibitem{ref36}
M. A. Richards, ``Fundamentals of Radar Signal Processing,'' New York, NY, USA: McGraw-Hill, 2014.

\bibitem{ref37}
L. Wang, Y. Zhang, Q. Liao, and J. Tang, ``Robust Waveform Design for Multi-target Detection in Cognitive MIMO Radar,'' in \textit{Proc. of 2018 IEEE Radar Conference (RadarConf18)},Oklahoma, USA, Apr. 2018, pp. 1--5.

\bibitem{ref38}
L. Wang, J. Tang, and Q. M. Liao, ``A Bayesian approach to target detection in cognitive radar,'' in \textit{Proc. of 2016 International Conference on Radar}, Oct. 2016.

\bibitem{ref39}
Z.-q. Luo, W.-k. Ma, A. M.-c. So, Y. Ye, and S. Zhang, `Semidefinite relaxation of quadratic optimization problems,'' \textit{IEEE Signal Process.}, vol. 27, no. 3, pp. 20--34, 2010.

\bibitem{ref40}
X. Yu, J. Chen, J. Zhang, and K. B. Letaief, `Alternating minimization algorithms for hybrid precoding in millimeter wave MIMO systems,'' \textit{IEEE J. Sel. Topics Signal Process.}, vol. 10, no. 3, pp. 485--500, Apr. 2016.

\bibitem{ref41}
Q. Wu, and R. Zhang, ``Intelligent reflecting surface enhanced wireless network: joint active and passive beamforming design,'' in \textit{Proc. of 2018 IEEE Global Communications Conference (GLOBECOM)}, Abu Dhabi, United Arab Emirates, Feb. 2019, pp. 1--6.

\end{thebibliography}
\end{document}